\documentclass[%
superscriptaddress,
preprint,
footinbib,
nobibnotes,
 amsmath,amssymb,
 aps,
]{revtex4-2}

\usepackage{hyperref}
\hypersetup{
    colorlinks=false,
    linkcolor=red,
    filecolor=black,      
    urlcolor=blue,
    citecolor=green
}

\usepackage[font=small]{caption}
\usepackage[labelformat=simple]{subcaption}

\usepackage{graphicx}
\usepackage{dcolumn}
\usepackage{bm}
\usepackage{footnote}
\DeclareMathOperator{\im}{i}
\DeclareMathOperator{\Tr}{Tr}
\def\Arg{\mathop{\operator@font arg}\nolimits}

\providecommand{\norm}[1]{\lVert#1\rVert}

\usepackage{color}

\begin{document}


\title{Mean-field model for a mixture of biaxial nematogens and dipolar nanoparticles}

\author{William G. C. Oropesa}
\email{carreras@if.usp.br}
\affiliation{%
Universidade de Sao Paulo, Instituto de Fisica, Rua do Matao, 1371, 05508-090, São Paulo, SP, Brazil\\
}
\affiliation{%
ICTP South American Institute for Fundamental Research,
Instituto de Física Teórica, UNESP — Univ. Estadual Paulista,
Rua Dr. Bento Teobaldo Ferraz 271, 01140-070, São Paulo, SP, Brazil\\
}

\author{Eduardo S. Nascimento}%
 \email{edusantos18@esp.puc-rio.br}
 \affiliation{%
Dept. of Physics, PUC-Rio, Rua Marquês de São Vicente 225,
22453-900 Rio de Janeiro, Rio de Janeiro, Brazi\\
}
 \author{André P. Vieira}%
 \email{apvieira@if.usp.br}
\affiliation{%
Universidade de São Paulo, Instituto de Fisica, Rua do Matao, 1371, 05508-090, São Paulo, SP, Brazil\\
}%

\date{\today}

\begin{abstract}
We analyze a mean-field model for mixtures involving biaxial nematogens and dipolar nanoparticles, taking into account orientational and isotropic pair interactions between nematogens, but also orientational nematogen-nanoparticle interactions. We determine bulk equilibrium phase diagrams for a wide range of interaction strengths, identifying in each case the effect of the nanoparticles on  the stability of nematic phases and on the appearance of multicritical points. Special attention is given to the limit of low concentration of nanoparticles, in which their effect on the temperatures of both the first-order uniaxial-isotropic and the continuous biaxial-uniaxial transitions is investigated in detail. 
\end{abstract}

\maketitle
\newpage

\section{Introduction\protect}\label{sec:level1}

Over the past few decades, the challenge of enhancing the physical properties of liquid crystals (LCs) through physical means, rather than chemical synthesis, has emerged. One approach to address this challenge is by incorporating colloidal particles into the LC host. The effects induced by these colloidal particles are heavily influenced by their size. Larger micrometric particles cause elastic distortions in the LC host, leading to an indirect interaction between them. Consequently, these microparticles can self-organize into a periodic array with potential applications in photonics~\cite{poulin1997novel, gu2000observation, stark2001physics, yada2004direct, smalyukh2005elasticity, musevic2006two}.

On the other hand, nanoparticles (NPs) are significantly smaller and do not cause distortion in the LC host. However, their addition can still produce substantial changes in the effective properties of the LC, as
shown by different experiments reporting both increases and decreases in the clearing temperature (the temperature at which the LC phase transitions into an isotropic liquid) when using various methods~\cite{li2006orientational,lin2015phase, kurochkin2010nano, ouskova2003dielectric, hakobyan2014enhanced, vcopivc2007coupled, huang2008probing, pinkevich1992nematic, buchnev2007enhanced, kaczmarek2008ferroelectric}; see also Refs. \cite{Prakash2020,Garbovskiy2017} and references therein.

Theoretical attempts to describe these experimental observations have involved Landau--de~Gennes expansions \cite{lopatina2009theory,lopatina2011maier}, molecular dynamics simulations \cite{Pereira2010}, or mean-field calculations \cite{gorkunov2011mean}, all of which focus on uniaxial nematogens. Here, we propose a theoretical approach which extends the analysis to deal with intrinsically biaxial nematogens, allowing us to also investigate the effects of NPs on the uniaxial-biaxial nematic phase transition.

In the two-tensor formulation of Sonnet, Virga, and Durand \cite{sonnet2003dielectric}, the anisotropic interaction potential $V^{\mathrm{nn}}_{ij}$ between two nematogens labeled as $i$ and $j$ can be written as
\begin{equation}
V^{\mathrm{nn}}_{ij} = -\frac{9}{4}A\left\{ \mathbf{q}_i\mathbf{:}\mathbf{q}_j + \zeta\left( \mathbf{q}_i\mathbf{:}\mathbf{b}_j + \mathbf{b}_i\mathbf{:}\mathbf{q}_j \right) + \lambda \mathbf{b}_i\mathbf{:}\mathbf{b}_j\right\}.
\label{eq:Vij}
\end{equation}
Here, $A>0$ sets the energy scale, while 
the second-rank tensors $\mathbf{q}$ and $\mathbf{b}$ are defined in terms of mutually orthogonal unit vectors 
$\hat{n}_1$, $\hat{n}_2$ and $\hat{n}_3$ pointing along the first, second and third principal axes of each nematogen as
\begin{equation}
\mathbf{q}=\hat{n}_{1}\otimes\hat{n}_{1}-\frac{1}{3}\mathbf{I}\quad\text{and}\quad\mathbf{b}=\hat{n}_{2}\otimes\hat{n}_{2}-\hat{n}_{3}\otimes\hat{n}_{3},
\end{equation}
$\mathbf{I}$ being the $3\times3$ identity matrix. 
The notation $\mathbf{q}_i\mathbf{:}\mathbf{q}_j$
stands for $\Tr\left(\mathbf{q}_i\mathbf{q}_j\right)$, where $\Tr\textbf{M}$ is the trace of matrix $\textbf{M}$.
Finally, the adimensional parameters $\zeta$ and $\lambda$ gauge the importance of biaxial couplings. Here
we work with the condition $\lambda=\zeta^2$, corresponding to the London approximation for dispersion forces \cite{sonnet2003dielectric,nascimento2016lattice,oropesa2022phase, oropesaphase}.

Within such formulation, in order to describe the anisotropic interaction potential $V^{\mathrm{nd}}_{ij}$ between a nematogen, labeled as $i$, and a uniaxial dipolar NP, labeled $j$, it is convenient to define the tensor
\begin{equation}
    \bm{\mathrm{d}}_{j}=\bm{\mathrm{p}}_{j}\otimes\bm{\mathrm{p}}_{j}-\dfrac{1}{3}\bm{\mathrm{I}},
\end{equation}
where the unit vector $\bm{\mathrm{p}}_{j}$ represents the dipolar moment of the NP. 
Then, $V^{\mathrm{nd}}_{ij}$ takes the form
\begin{equation}
    V^{\mathrm{nd}}_{ij}=-\dfrac{9}{4}B\left(\bm{\mathrm{q}}_{i}\bm{:}\bm{\mathrm{d}}_{j}+\xi\bm{\mathrm{b}}_{i}\bm{:}\bm{\mathrm{d}}_{j}\right)=-\dfrac{9}{4}B\left(\bm{\mathrm{q}}_{i}+\xi\bm{\mathrm{b}}_{i}\right)\bm{:}\bm{\mathrm{d}}_{j}.
\end{equation}
The dimensionless parameter $\xi$ gauges the importance of the coupling between the dipole and the nematogenic biaxial part. For $\xi=0$, we recover the Maier--Saupe interaction energy which is appropriate only for the description of intrinsically uniaxial nematogens. For intrinsically biaxial nematogens, we need to set the parameter $\xi$ to a nonzero value. 

In order to model a binary mixture of nematogens and dipolar NPs, we assume that the particles occupy the sites of a regular lattice, only single occupation being allowed. In this problem, similar to the dilution problem discussed in Refs.~\cite{oropesa2022phase, oropesaphase}, we also introduce an isotropic interaction $U$ between nematogens. Combining all previously discussed interactions, and assuming $\zeta=\xi\equiv\Delta/3$, where $\Delta$ gauges the intrinsic biaxiality of the nematogens, the contribution of two neighboring sites $i$ and $j$ to the total interaction energy of the system is
\begin{equation}\label{eq:lml}
\begin{split}
    V_{ij}&=\gamma_{i}\gamma_{j}\left[U-\dfrac{9}{4}A\left(\bm{\mathrm{q}}_{i}+\dfrac{\Delta}{3}\bm{\mathrm{b}}_{i}\right)\bm{:}\left(\bm{\mathrm{q}}_{j}+\dfrac{\Delta}{3}\bm{\mathrm{b}}_{j}\right)\right]\\
    &\qquad
    -\dfrac{9}{4}B\left[\gamma_{i}(1-\gamma_{j})\left(\bm{\mathrm{q}}_{i}+\dfrac{\Delta}{3}\bm{\mathrm{b}}_{i}\right)\bm{:}\bm{\mathrm{d}}_{j}
    +(1-\gamma_{i})\gamma_{j}\bm{\mathrm{d}}_{i}\bm{:}\left(\bm{\mathrm{q}}_{j}+\dfrac{\Delta}{3}\bm{\mathrm{b}}_{j}\right)\right].
\end{split}
\end{equation}
\noindent
The occupation variable $\gamma_{i}$ is equal to $0$ if site $i$ is occupied by a dipolar NP and to $1$ if the site is occupied by a nematogen. Here we allow the strength of the anisotropic interaction $B$ between objects of different nature to be either negative or positive. The case $B>0$ energetically favors configurations in which a dipole aligns (or anti-aligns) with the first principal axis of a nematogen. On the other hand, the case $B<0$ energetically favors configurations in which a dipole is perpendicular to the first principal axis of a nematogen. Finally, the case $B=0$ corresponds to an effective diluted problem, for which $\gamma_{i}=0$ represents a hole.

The use of lattice models to describe both isotropic and anisotropic fluids, either pure or in mixtures, has a long history in statistical mechanics; see e.g. Refs.\,\cite{Lee1952, lebwohl1972nematic, Vieira1999, bates2001computer} and references therein. Indeed, in Ref.~\cite{oropesa2022phase}, we presented a lattice-gas Maier--Saupe--Zwanzig (LGMSZ) model for the description of the effects of dilution in nematic LCs with intrinsically-biaxial nematogens. The model combines a mean-field version of the interaction potential in
Eq.\,(\ref{eq:lml}), for the case $B=0$, with the Zwanzig approximation \cite{zwanzig1963first}, which restricts the possible orientations of a nematogen to the coordinate axes.
In the present paper, the holes are exchanged for dipolar NPs that do not directly interact with each other, but do interact with the nematogens. As the strength of the interaction between nematogens and dipolar NPs is $B$, and the orientations of the dipolar NPs are also subject to the Zwanzig approximation, the lattice model used to describe the binary mixture will be called the LGMSZ-B model.

It is worth mentioning that the Zwanzig approximation was originally proposed \cite{zwanzig1963first} to deal with a model of long thin hard rods for the nematic transition, and has been later adopted by a number of authors \cite{Shih1972, Chen1984, de1986reentrant, Taylor1991, Henriques1997, Belli2011, Reinink2014, sauerwein2016lattice}. Although it may not be adequate to account for the properties of off-lattice systems with hard boardlike particles \cite{Straley1972, Shundyak2004, Patti2018}, it has been used to obtain a number of equilibrium features of  nematic systems at the mean-field level \cite{de1986reentrant, do2010statistical, do2011phase, liarte2012enhancement, nascimento2015maier, sauerwein2016lattice, nascimento2016lattice, petri2018field, rodrigues2020magnetic, dosSantos2021, DeMatteis2023}, always leading to qualitative results which fully agree with continuous versions of the corresponding models when a comparison is possible. In particular, when dealing with intrinsically biaxial nematogens, these models are capable of reproducing the qualitative characteristics of nematic phase diagrams, such as sequences of biaxial-uniaxial-isotropic phase transitions with increasing temperature, and a well-defined Landau multicritical point, which signals a direct transition between the isotropic and the biaxial phases \cite{boccara1977solvable,nascimento2016lattice,dosSantos2021}.

We note that, in principle, neglecting interactions between dipolar nanoparticles is a reasonable approximation only in the limit where the relative concentration of such particles is small. However, in experimental systems there are also ionic impurities which may screen electrostatic interactions (see Ref. \cite{Garbovskiy2015} and references therein), therefore extending the validity of the approximation to higher concentrations, while at the same time weakening the coupling between NPs and nematogens. For the sake of completeness, we extend the approximation to the full range of possible concentrations, with the warning that results for the highest values should be subject to further checks. 
Within this full range of concentrations, as in other investigations focusing on different binary mixtures of fluids \cite{Konynenburg1980, Chen1984, Henriques1997, henriques2008mixture, do2011phase, nascimento2015maier, Sidky2016, oropesa2022phase}, we encounter multiple instances of phase coexistence, which we describe in detail.

This paper is organized as follows. Sec.~\ref{sec:lgmszb} presents the LGMSZ-B model, summarizes its mean-field solution, and discusses the stability of multicritical Landau points for different values of the degree of biaxiality of nematogens. In Sec.~\ref{sec:u0} we present an analysis of the effects of anisotropic interactions between dipolar NPs and nematogens with fixed degree of biaxiality, in the case of zero isotropic interaction. Sec.~\ref{sec:lcbs} is dedicated to LC-based suspensions and the effects of NPs on the clearing temperature as well as on the uniaxial-biaxial second-order transition temperature. In Sec.~\ref{sec:unoo} we present a study of the effects of the isotropic interaction for fixed biaxiality degree and anisotropic interaction between objects of different nature. Conclusions are presented in Sec.~\ref{sec:conp}. A few technical details are relegated to Appendices~\ref{app:A},~\ref{app:B}, and~\ref{app:C}.

\section{The LGMSZ-B model\protect}\label{sec:lgmszb}
We consider a system which is a binary mixture consisting of $N_{m}$ instrinsically biaxial nematogens and $N_{d}$ uniaxial dipolar NPs. Each lattice site can be occupied by a nematogen or by a dipole, the state of site $i$ being described by an occupation variable $\gamma_{i}$ taking the values 0 (dipole) or 1 (nematogen). 
Then, for an arbitrary lattice, we define the LGMSZ-B model with dipolar-quadrupolar couplings by means of the effective Hamiltonian
\begin{equation}\label{eq:model}
    \begin{split}
    \mathcal{H} = \sum_{(i,j)}V_{ij} &= U\sum_{(i,j)}\gamma_i\gamma_j 
    -A\sum_{(i,j)}\gamma_j\gamma_j\bm{\Omega}_{i}\bm{:}\bm{\Omega}_{j} \\
    &-B\sum_{(i,j)}\left[\gamma_i(1-\gamma_j)\bm{\Omega}_{i}\bm{:}\bm{\Theta}_{j}
                       +(1-\gamma_i)\gamma_j\bm{\Theta}_{i}\bm{:}\bm{\Omega}_{j}\right],
    \end{split}
\end{equation}
where $A$, $B$ and $U$ are coupling parameters, with $A>0$, the sum is performed over pairs $(i,j)$ of neighboring sites $i$ and $j$ in the lattice, and the quantities $\bm{\Omega}_{i}\equiv \frac{3}{2}\left(\bm{\mathrm{q}}_{i}+\frac{\Delta}{3}\bm{\mathrm{b}}_{i}\right)$ and $\bm{\Theta}_{j}=\frac{3}{2}\bm{\mathrm{d}}_{j}$ are second-rank tensors associated with a nematogen at site $i$ and a dipolar NP at site $j$, respectively. 
By resorting to a simplified view of a biaxial nematogen as a rectangular platelet, the biaxiality parameter $\Delta$ can be interpreted in terms of the sides of the platelet, so that $\Delta=0$ corresponds to a ``rod-like'' object, $\Delta=3$ to a ``disk-like'' object and $\Delta=1$ to a maximally biaxial object \cite{nascimento2015maier}. In this paper, we restrict ourselves to the cases $0\leqslant\Delta\leqslant3$.

Furthermore, as already discussed in the Introduction, instead of working with continuous orientational states, we follow the Zwanzig prescription in assuming that the principal axes of a nematogen and the dipolar axes are restricted to align in the directions of the Cartesian axes.
Note that, for zero isotropic interaction ($U=0$) and intrinsically uniaxial molecules ($\Delta=0$), Eq.\,(\ref{eq:model}) reduces to a discretized version of the continuous model introduced by A. N. Zakhlevnykh \textit{et al.}~\cite{zakhlevnykh2016simple} for suspensions of anisometric particles in nematic liquid crystals (with a rescaling of energies, as our parameters $A$ and $B$ would be respectively equivalent to $\sqrt{3/2}A$ and $\sqrt{3/2}B$ in the units of Ref.~\cite{zakhlevnykh2016simple}).

In order to make analytic progress in the problem defined by Eq.\,(\ref{eq:model}), we resort to a mean-field (MF) treatment, which is equivalent
to considering the fully-connected Hamiltonian
\begin{equation}
    \mathcal{H}_{\mathrm{mf}}=-\dfrac{A}{2N}\sum^{N}_{i,j=1}\gamma_{i}\gamma_{j}\bm{\Omega}_{i}\bm{:}\bm{\Omega}_{j}-\dfrac{B}{N}\sum^{N}_{i,j=1}\gamma_i(1-\gamma_j)\bm{\Omega}_{i}\bm{:}\bm{\Theta}_{j}+\dfrac{U}{2N}\sum^{N}_{i,j=1}\gamma_{i}\gamma_{j},
\end{equation}
where the sums over pairs of neighboring sites are replaced by sums over all pairs of sites,
and factors of $1/N$ are included to ensure that energy is extensive. In this problem, $N=N_m+N_d$ is the number of objects (nematogens and dipolar NPs) of the binary mixture. It is then convenient to consider
the formalism of the grand canonical ensemble, where the numbers of nematogens and of dipolar NPs may fluctuate due to the coupling to a particle reservoir~\cite{nascimento2015maier, do2010statistical, rodrigues2020magnetic}. Thus, we must determine the grand partition function
\begin{equation}
    \Xi=\sum_{\{\gamma_{i}\}} \sum_{\{\bm{\Omega}_{i}\}} \sum_{\{\bm{\Theta}_{i}\}} \exp\left(\beta\mathcal{H}_{\mathrm{mf}}-\beta\mu\sum^{N}_{i=1}\gamma_{i}\right),
\end{equation}
where $\beta=1/k_BT$, $k_B$ is the Boltzmann constant (which we take to be equal to 1 in suitable units), $T$ is the temperature and $\mu$ is the chemical potential, which controls the number of nematogens. 
All quantities with dimension of energy ($T$, $\mu$, $B$, and $U$) are measured in units of $A$.

The MF calculations, detailed in Appendix~\ref{app:A}, yield a Landau--de Gennes free-energy functional $\psi\left(S,R,\eta,\zeta,\phi\right)$. The equilibrium values of the scalar variables $S$, $\eta$, $R$, $\zeta$, and $\phi$, i.e. those which correspond to absolute minima of $\psi$, are associated with the ensemble averages
\begin{equation} 
    \left< \frac{1}{N}\sum_{i=1}^N\gamma_i\bm{\Omega_i} \right> = \frac{1}{2}
    \begin{pmatrix}
    -S - \eta & 0 & 0 \\
    0 & -S + \eta & 0 \\
    0 & 0 &  2S
    \end{pmatrix},
\end{equation}
\begin{equation} 
    \left< \frac{1}{N}\sum_{i=1}^N(1-\gamma_i)\bm{\Theta_i} \right> = \frac{1}{2}
    \begin{pmatrix}
    -R - \zeta & 0 & 0 \\
    0 & -R + \zeta & 0 \\
    0 & 0 &  2R
    \end{pmatrix},
\end{equation}
and
\begin{equation}
\left< \frac{1}{N}\sum_{i=1}^N\gamma_i \right> = \phi.
\end{equation}
Therefore, at thermal equilibrium, $S$ and $\eta$ measure the uniaxial and biaxial ordering of the nematogens, $R$ and $\zeta$ play analogous roles for the dipolar NPs, and $\phi$ corresponds to the concentration of nematogens. 
Table\,\ref{tab:I} summarizes the symbols used in this paper.
\begin{table}[]
\begin{tabular}{ll}
\hline
$A$ & strength of the orientational coupling between two quadrupolar particles \\ \hline
$B$ & strength of the orientational coupling between a quadrupolar and a dipolar particle \\ \hline
$U$ & strength of the isotropic coupling between two quadrupolar particles \\ \hline
$T$ & temperature (Boltzmann's constant $k_B=1$) \\ \hline
$\beta$ & inverse temperature \\ \hline
$\mu$ & chemical potential \\ \hline
$\Delta$ & biaxiality parameter of quadrupolar particles \\ \hline
$\kappa$ & ratio between $B$ and $A$ \\ \hline
$\phi$ & relative concentration of quadrupolar particles \\ \hline
$S$ & uniaxial order parameter of the quadrupolar particles \\ \hline
$\eta$ & biaxial order parameter of the quadrupolar particles \\ \hline
$R$ & uniaxial order parameter of the dipolar particles \\ \hline
$\zeta$ & biaxial order parameter of the dipolar particles \\ \hline
\end{tabular}
\caption{
List of symbols used throughout the text.
}
\label{tab:I}
\end{table}

Explicitly, we have
\begin{equation}
\begin{split}
    \psi(S,R,\eta,\zeta,\phi)&=\dfrac{A}{4}(3S^{2}+\eta^{2})+\dfrac{U}{2}\phi^{2}-\mu\phi+\dfrac{\zeta-R}{3\beta}\ln\left[(R-1+\phi)^{2}-\zeta^{2}\right]\\
    &\qquad + \dfrac{1-\phi}{3\beta}\ln\left\{(1+2R-\phi)\left[(R-1+\phi)^{2}-\zeta^{2}\right]\right\}\\
    &\qquad +\dfrac{2R}{3\beta}\ln\left(1+2R-\phi\right)-\dfrac{2\zeta}{3\beta}\ln\left(1-R-\zeta-\phi\right)\\
    &\qquad -\dfrac{\phi}{\beta}\ln\left[\Lambda(S,R,\eta,\zeta)\right]+\dfrac{\phi}{\beta}\ln(6\phi)-\dfrac{2}{\beta}\ln(6),
\end{split}
\end{equation}
where
\begin{equation}
  \begin{split}
\Lambda(S, R, \eta, \zeta) &= 2\exp\left\{ -\frac{3\beta}{4}\left[A(S+\eta)+B(R+\zeta) \right]\right\} \\
&\quad\times\cosh{\left\{\dfrac{3\beta}{4}\left[A\left(S-\dfrac{\eta}{3}\right)+B\left(R-\dfrac{\zeta}{3}\right)\right]\Delta\right\}} \\
&\quad+ 2\exp\left\{ -\frac{3\beta}{4}\left[A(S-\eta)+B(R-\zeta) \right]\right\} \\
&\quad\times\cosh{\left\{\dfrac{3\beta}{4}\left[A\left(S+\dfrac{\eta}{3}\right)+B\left(R+\dfrac{\zeta}{3}\right)\right]\Delta\right\}} \\
	&\quad + 2\exp\left[\frac{3\beta}{2}(AS+BR)\right] \cosh{\left[\dfrac{\beta}{2}(A\eta+B\zeta) \Delta\right]}.
\end{split}
\end{equation}

The equilibrium values of $S$, $R$, $\eta$, $\zeta$ and $\phi$ are determined by locating the absolute minima of $\psi(S,R,\eta,\zeta,\phi)$, leading to the mean-field (MF) equations
\begin{equation}\label{eq:mfe}
    \dfrac{\partial\psi}{\partial S}=\dfrac{\partial \psi}{\partial \eta} = \dfrac{\partial \psi}{\partial R}=\dfrac{\partial\psi}{\partial \zeta}=\dfrac{\partial\psi}{\partial\phi} = 0,
\end{equation}
which take the self-consistent forms 
\begin{equation}
\begin{matrix}
S=F_{1}(S,R,\eta,\zeta,\phi;\{\alpha\})\\ 
\eta=F_{2}(S,R,\eta,\zeta,\phi;\{\alpha\})\\ 
R=F_{3}(S,R,\eta,\zeta,\phi;\{\alpha\})\\ 
\zeta=F_{4}(S,R,\eta,\zeta,\phi;\{\alpha\})\\ 
\phi=F_{5}(S,R,\eta,\zeta,\phi;\{\alpha\})
\end{matrix}
\end{equation}
where $\{\alpha\}\equiv\{\beta,\mu,\Delta, A, B\}$ is a set of parameters. The free energy $\mathcal{F}=\mathcal{F}(\{\alpha\})$ of the system corresponds to the convex envelope of $\psi$ determined after inserting values of $S$, $\eta$, $R$, $\zeta$ and $\phi$ associated with the minima of the Landau--de Gennes free-energy functional.

In order to obtain equilibrium phase diagrams, we must study how the equilibrium values of $S$, $R$, $\eta$, $\zeta$ and $\phi$ change as the parameters in $\{\alpha\}$ are varied. Depending on these parameters, the system may exhibit uniaxial and biaxial nematic phases, at sufficiently low temperature and sufficiently high nematogen concentration, as well as an isotropic fluid phase. 
For a fixed $\{\alpha\}$, the isotropic phase has only one equilibrium solution, characterized by $S=\eta=R=\zeta=0$. There are two uniaxial nematic phases, called calamitic and discotic, in which our model has 3 equilibrium solutions.
One of the solutions in the calamitic (resp. discotic) phase has $\eta=\zeta=0$ and $S>0$ (resp. $S<0$), while the sign of $R$ depends on the interaction parameter $B$. Finally, the biaxial nematic phase has 6 equilibrium solutions, all of which generically have nonzero values of $S$, $\eta$, $R$ and $\zeta$. There is also the possibility of phase coexistence, as discussed below. The reader should keep in mind that the finite number of equilibrium solutions in the ordered phases is an artifact of the Zwanzig approximation, and a continuous model would instead have an infinite number of solutions.

The recipes for locating the various transition lines and multicritical points are outlined in Secs.\,\ref{sec:u0} and \ref{sec:unoo}. A discussion on the Landau multicritical point, at which the isotropic phase and various nematic phases become identical, can be found in Appendix \ref{sec:lp}.

\begin{figure}[t]
    \begin{minipage}[b]{0.325\textwidth}
    \centering
     \includegraphics[width=\textwidth]{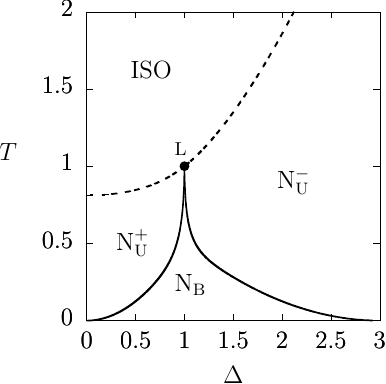}
     \subcaption{}
     \label{fig:TvsDuni}
     \end{minipage}
     \hspace{10pt}
    \begin{minipage}[b]{0.3\textwidth}
     \centering
     \includegraphics[width=\textwidth]{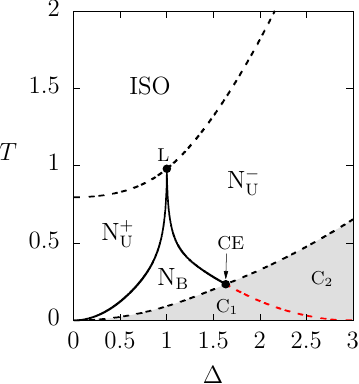}
     \subcaption{}
     \label{fig:TvsDBp}
    \end{minipage}
    \hspace{10pt}
    \begin{minipage}[b]{0.3\textwidth}
     \centering
     \includegraphics[width=\textwidth]{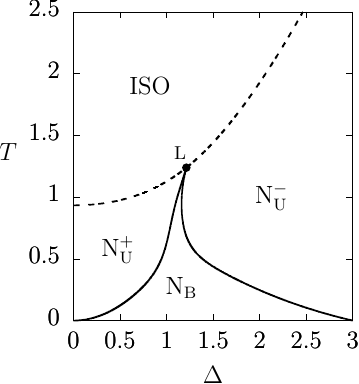}
     \subcaption{}
     \label{fig:TvsDBp2}
    \end{minipage}
    
    \begin{center}
    \begin{minipage}[b]{0.3\textwidth}
     \centering
     \includegraphics[width=\textwidth]{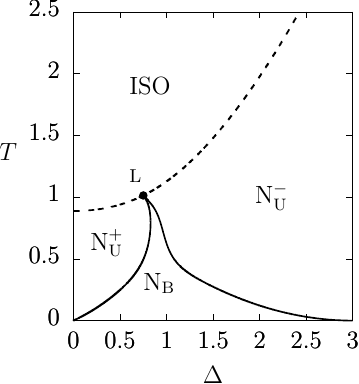}
     \subcaption{}
     \label{fig:TvsDBn2}
    \end{minipage}
    \hspace{10pt}
    \begin{minipage}[b]{0.3\textwidth}
     \centering
     \includegraphics[width=\textwidth]{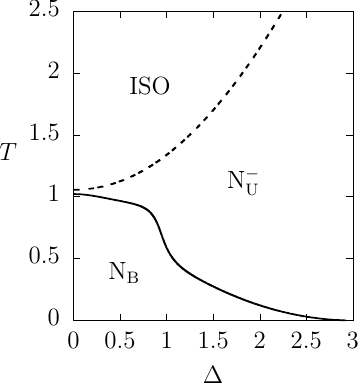}
     \subcaption{}
     \label{fig:TvsDBn}
    \end{minipage}
    \end{center}
    
    \caption{
    Temperature versus nematogen shape biaxiality $\Delta$, for $U=0$. (a) Pure system ($\phi=1$). (b) Dipole-doped system, with $\phi=0.975$ and $B=A/2$. (c) Dipole-doped system, with $\phi=0.975$ and $B=3A$. (d) Dipole-doped system, with $\phi=0.975$ and $B=-3A$. (e) Dipole-doped system, with $\phi=0.975$ and $B=-5A$. $\mathrm{ISO}$: isotropic phase. $\mathrm{N^{+}_{U}}$: calamitic uniaxial nematic phase. $\mathrm{N^{-}_{U}}$: discotic uniaxial nematic phase. $\mathrm{N_{B}}$: biaxial nematic phase. $\mathrm{C}_1$: coexistence between $\mathrm{ISO}$ and $\mathrm{N_{B}}$. $\mathrm{C}_2$: coexistence between $\mathrm{ISO}$ and $\mathrm{N^{-}_{U}}$. $\mathrm{L}$ is a Landau multicritical point. $\mathrm{CE}$ in (b) is a critical end point (see Sec.\,\ref{sec:u0}) and the contiguous dashed red (dark gray) line is a line of CE points separating $\mathrm{C}_1$ and $\mathrm{C}_2$. }
    \label{fig:TvsD}
\end{figure}
Our main focus in this paper are phase diagrams in the plane of temperature $T$ versus nematogen concentration $\phi$. Such diagrams will be more systematically explored starting in the next Section. Nevertheless, in the remainder of the current Section, we offer a comparison with the behavior in the pure limit $\phi=1$, by drawing in Fig.\,\ref{fig:TvsD} some diagrams in the plane of temperature versus biaxiality degree $\Delta$, for systems in which nematogen concentration corresponds to $\phi=97.5\%$, with no isotropic interaction, i.e. $U=0$.

Figure\,\ref{fig:TvsDuni} shows the well-known $T$ versus $\Delta$ phase diagram in the pure limit $\phi=1$ 
\cite{boccara1977solvable,
Biscarini1995,
Berardi2008,
nascimento2015maier, 
nascimento2016lattice,
rodrigues2020magnetic},
exhibiting two lines of continuous transitions
(solid lines) from a biaxial nematic region ($\mathrm{N_B}$) to the calamitic ($\mathrm{N_U^+}$)
and the discotic ($\mathrm{N_U^-}$) uniaxial nematic regions. These critical lines
meet at a Landau multicritical point (L) on the first-order
boundary (dashed lines) between the isotropic and the uniaxial
nematic phases. With our choice of units and parameters, the Landau point is located at $T_\mathrm{L}=A$ and $\Delta_\mathrm{L}=1$, corresponding to maximal biaxiality (a feature present even for short-range interactions \cite{Biscarini1995}). When temperature is lowered in systems with $\Delta=1$, the stable phase changes directly from an isotropic fluid to a biaxial nematic phase, without the intermediate uniaxial nematic phase observed for $0<\Delta<1$ or $1<\Delta<3$.

Adding a small amount of dipolar particles, thereby reducing the nematogen concentration to $\phi=0.975$, changes the position of the Landau point and, depending on the value of the coupling $B$ between nematogens and dipolar particles, may also change the topology of the phase diagram, as illustrated in Figs.\,\ref{fig:TvsDBp} through  \ref{fig:TvsDBn}. For $B=0$, as discussed in Ref.\,\cite{rodrigues2020magnetic}, the temperature of the Landau point is reduced but the point is still located along the line $\Delta=1$; see also Appendix\,\ref{sec:lp}. On the other hand, for $B\neq 0$, the Landau point is shifted away from maximal nematogen biaxiality, with $\Delta_\mathrm{L}>1$ if $B>0$ and $\Delta_\mathrm{L}<1$ if $B<0$. This shift can be qualitatively understood by noticing that $B>0$ (resp. $B<0$) favors a parallel (resp. perpendicular) alignment between a dipole and the first principal axis of a nematogen, inducing a kind of effective average particle shape which is more cylindrical (resp. discotic) and requiring a larger (resp. smaller) value of $\Delta$ to achieve maximal effective biaxiality. A sufficiently large but negative $B$ induces the disappearance of the Landau point, as illustrated in Fig.\,\ref{fig:TvsDBn} for $B=-5A$, due to the fact that, with our energy parametrization, in this limit there is no stable calamitic phase even at very low temperatures.

Away from the Landau point, a small $B$ induces other changes in the phase diagrams, as illustrated in Fig.\,\ref{fig:TvsDBp} for $B=A/2$. The shaded regions indicate coexistence between a dipolar-rich isotropic phase and a nematogen-rich ordered phase which is either biaxial (region marked as $\mathrm{C}_1$) or discotic (region marked as $\mathrm{C}_2$). The small fraction of the isotropic phase and the large fraction of the ordered phase in coexistence combine to yield an average concentration $\phi=0.975$. These phase coexistences are further discussed in the next Section. They are absent for larger values of $B$ due to the low-temperature phase behavior discussed in Appendix\,\ref{app:B}. The phase diagrams for small but negative $B$ look like distorted mirror images of Fig.\,\ref{fig:TvsDBp}, with the calamitic phase replacing the discotic phase in region $\mathrm{C}_2$. 

Other topologies of the $T$ versus $\Delta$ phase diagram would appear for smaller nematogen concentration, as can be inferred from the $T$ versus $\phi$ phase diagrams shown in the next Sections. For the sake of space, we will not discuss these other topologies here.

\section{Phase behavior for \texorpdfstring{$U=0$}{}}\label{sec:u0}
\begin{figure}[t]
    \centering
    \begin{minipage}[b]{0.48\textwidth}
     \centering
     \includegraphics[scale=1]{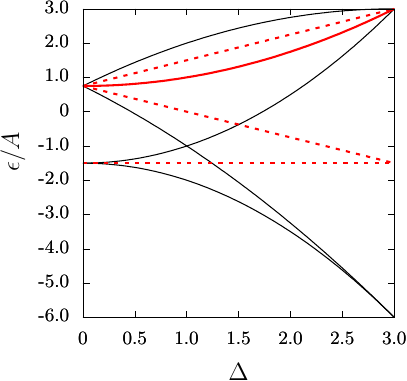}
     \subcaption{$B=1$}
     \label{fig:232a}
    \end{minipage}
    \hspace{10pt}
    \begin{minipage}[b]{0.48\textwidth}
     \centering
     \includegraphics[scale=1]{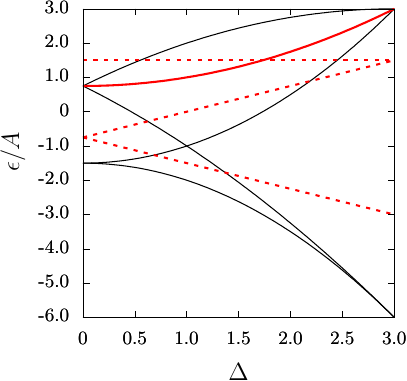}
     \subcaption{$B=-1$}
     \label{fig:232b}
    \end{minipage}
    \hspace{10pt}
    \caption{Energy levels as a function of the degree of biaxiality $\Delta$. Level crossings account for variations in the entropic contributions to the free energy. Solid lines correspond to energy levels associated with interactions between two nematogens, while dashed lines indicate energy levels arising from interactions involving NPs and nematogens. Line thickness is proportional to the degeneracy of the corresponding level.}
    \label{fig:232}
\end{figure}

For this binary system, the $T$ versus $\phi$ phase diagrams present a great variety of topologies. This is intuitive when observing the energy levels associated with two nematogens or a nematogen and a dipolar NP, plotted in Fig.\,\ref{fig:232}. Depending on the degree of biaxiality $\Delta$ of the nematogens, we can observe how energy levels with different degeneracies [12 for red (gray) curves and 6 for black curves] intersect. The degeneracies are associated with entropic contributions to the free energy and can produce important effects on the phase behavior of the system, mainly for values of $\Delta$ near the crossings points of energy levels. Another evidence of the large number of topologies that can be found is the dependence of the positions of the crossing points on the strength of the interaction between dipolar NPs and nematogens; see the differences between Figs.~\ref{fig:232a} and \ref{fig:232b}.
In this section, we start the investigation of the phase behavior for zero isotropic interaction ($U=0$), which simplifies the  analysis of the system. Here we discuss both positive and negative values for the strength of the nematogen-NP interaction $B$, as well as the effects of introducing small amounts of dipolar NPs into the LC host. 

When the phase diagrams presented have more than one uniaxial phase of the same type, either calamitic ($+$) or discotic ($-$), we will use Roman numerals to distinguish the phases, e.g. $\mathrm{N^{+}_{U_{I}}}$, $\mathrm{N^{+}_{U_{II}}}$, $\mathrm{N^{+}_{U_{III}}}$ etc.

\subsection{Phase diagrams for \texorpdfstring{$\Delta=0$}{}}\label{sec:pdd0}
\begin{figure}[ht]
    \centering
    \begin{minipage}[b]{0.45\textwidth}
     \centering
     \includegraphics[scale=0.9]{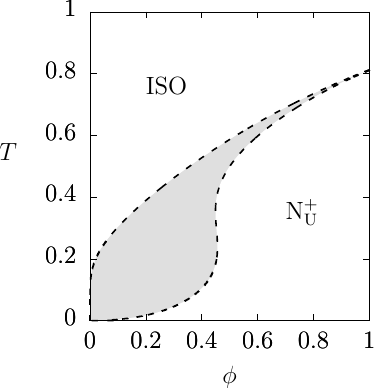}
     \subcaption{$(A, B)=(1,3/4)$}
     \label{fig:14a}
    \end{minipage}
    \hspace{10pt}
    \begin{minipage}[b]{0.45\textwidth}
     \centering
     \includegraphics[scale=0.9]{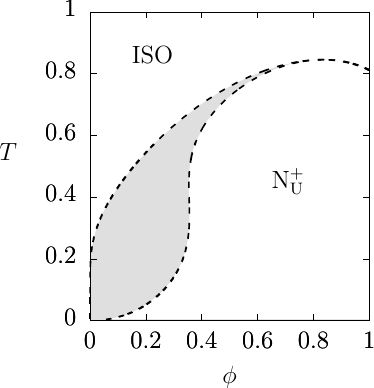}
     \subcaption{$(A, B)=(1,5/4)$}
     \label{fig:14b}
    \end{minipage}
    \caption{Phase diagram in terms of temperature $T$ (in units of $A$) and concentration $\phi$ of nematogens, for an intrinsically uniaxial system ($\Delta=0$), different values of anisotropic interaction between nematogens and dipolar NPs and in absence of isotropic interaction ($U=0$). $\mathrm{ISO}$: isotropic phase. $\mathrm{N^{+}_{U}}$: calamitic uniaxial nematic phase. Short-dashed lines are the boundaries of the biphasic region (gray).}
    \label{fig:14}
\end{figure}

We find the $\phi$-$T$ phase diagrams shown in Fig.\,\ref{fig:14}, for different positive values of the strength of the anisotropic interaction between dipolar nanoparticles and nematogens. For high concentration of nematogens, as the temperature decreases, the observed phase sequence in these diagrams is $\mathrm{ISO}$, followed by a narrow $\mathrm{ISO}$-$\mathrm{N^{+}_U}$ coexistence region, then by a pure nematogen-rich (or dipole-poor) $\mathrm{N^{+}_U}$ phase. At lower concentration of nematogens, as the temperature decreases, the observed phase sequence in the diagrams is $\mathrm{ISO}$, followed by an $\mathrm{ISO}$-$\mathrm{N^{+}_U}$ coexistence region, then by a pure nematogen-poor (or dipole-rich) low temperature $\mathrm{N^{+}_U}$ phase. The coexistence lines signaling the first-order transition from the $\mathrm{ISO}$ phase to the $\mathrm{N^{+}_U}$ phase are determined by Eq.\,(\ref{eq:mfe}) evaluated at $(S,\eta,R,\zeta, \phi)=(S_{\mathrm{U}}, 0, R_{\mathrm{U}}, 0, \phi_{\mathrm{U}})$ and at $(S,\eta,R,\zeta, \phi)=(0,0,0,0,\phi_{\mathrm{I}})$, supplemented by $\psi(S_{\mathrm{U}}, 0, R_{\mathrm{U}}, 0, \phi_{\mathrm{U}})=\psi(0,0,0,0,\phi_{\mathrm{I}})$, where $\phi_{\mathrm{I}}$ and $\phi_{\mathrm{U}}$ are, respectively, the nematogen concentration in the $\mathrm{ISO}$ and $\mathrm{N^{+}_U}$ phases at the transition point, while $S_{\mathrm{U}}$ and $R_{\mathrm{U}}$ are the values of $S$ and $R$ at that point. In this case, without loss of generality, we can assume $\eta=\zeta=0$.

\begin{figure}[ht]
    \centering
    \begin{minipage}[b]{0.475\textwidth}
     \centering
     \includegraphics[scale=0.9]{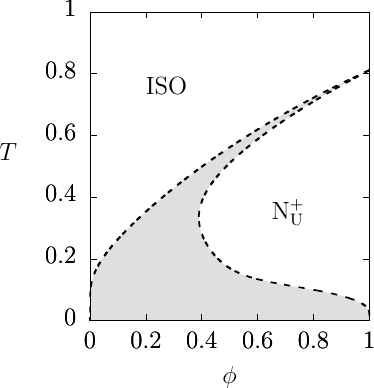}
     \subcaption{$(A, B)=(1,-3/4)$}
     \label{fig:24a}
    \end{minipage}
    \hspace{10pt}
    \begin{minipage}[b]{0.475\textwidth}
     \centering
     \includegraphics[scale=0.9]{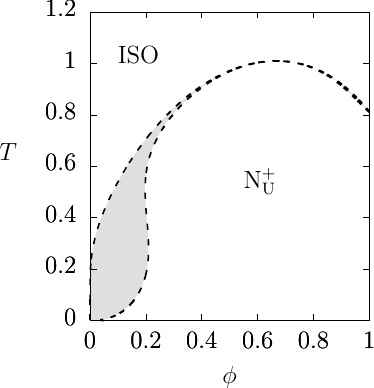}
     \subcaption{$(A, B)=(1,-2)$}
     \label{fig:24b}
    \end{minipage}
    \hspace{10pt}
    \caption{Phase diagram in terms of temperature $T$ (in units of $A$) and concentration $\phi$ of nematogens, for an intrinsically uniaxial system ($\Delta=0$), different values of anisotropic interaction between nematogens and dipolar NPs and in absence of isotropic interaction ($U=0$). $\mathrm{ISO}$: isotropic phase. $\mathrm{N^{+}_{U}}$: calamitic uniaxial nematic phase. Short-dashed lines are the boundaries of biphasic region (gray).}
    \label{fig:24}
\end{figure}

The phase diagrams for systems with negative interaction between dipolar NPs and nematogens ($B<0$) are shown in Fig.\,\ref{fig:24}. The $\phi$-$T$ phase diagram shown in Fig.\,\ref{fig:24a} corresponds to a system with $(A, B)=(1,-3/4)$. At lower concentrations of nematogens, as the temperature decreases, we observe an $\mathrm{ISO}$ phase, followed by an $\mathrm{ISO}$-$\mathrm{N^{+}_U}$ coexistence region stable at low temperature. At high concentration of nematogens, we observe an $\mathrm{ISO}$ phase, followed by a narrow $\mathrm{ISO}$-$\mathrm{N^{+}_U}$ coexistence region, then by a pure $\mathrm{N^{+}_{U}}$ phase and finally by a reentrant coexistence region. For $T\to0$, we have a stable $\mathrm{ISO}$-$\mathrm{N^{+}_U}$ biphasic region, which becomes metastable with respect to the $\mathrm{N^{+}_U}$ phase if $B<-A$ or $B>A/2$, as shown in Appendix \ref{app:B}.
Fig.\,\ref{fig:24b} represents a situation with strong anisotropic interaction ($B=-2A$) between dipolar NPs and nematogens. In the diagram, there is a large stability region for the $\mathrm{N^{+}_{U}}$ phase and the $\mathrm{ISO}$-$\mathrm{N^{+}_U}$ coexistence disappears as $T\rightarrow 0$. Note that, for high concentration of nematogens, the addition of dipolar NPs increases the temperature of the first-order $\mathrm{ISO}$-$\mathrm{N^{+}_U}$ phase transition.

\subsection{Phase diagrams for \texorpdfstring{$\Delta=4/5$}{}}\label{sec:pdd08}
\begin{figure}[ht]
    \centering
    \begin{minipage}[b]{0.475\textwidth}
     \centering
     \includegraphics[scale=0.9]{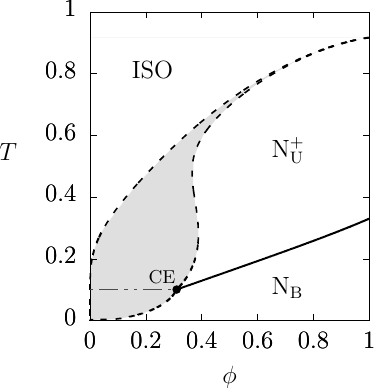}
     \subcaption{$(A, B)=(1,1)$}
     \label{fig:434a}
    \end{minipage}
    \hspace{10pt}
    \begin{minipage}[b]{0.475\textwidth}
     \centering
     \includegraphics[scale=0.9]{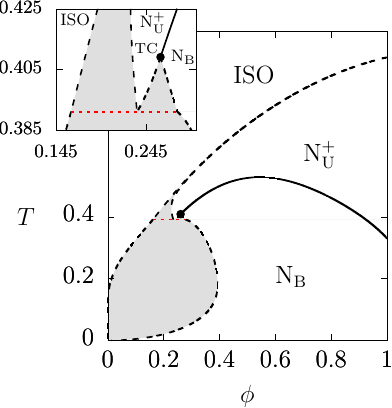}
     \subcaption{$(A, B)=(1,-1)$}
     \label{fig:434b}
    \end{minipage}
   \hspace{10pt}
    \begin{minipage}[b]{0.475\textwidth}
     \centering
     \includegraphics[scale=0.9]{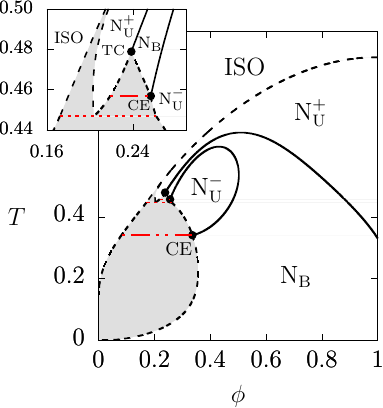}
     \subcaption{$(A, B)=(1,-23/20)$}
     \label{fig:434c}
    \end{minipage}
    \caption{Phase diagrams in terms of temperature $T$ (in units of $A$) and concentration $\phi$ of nematogens, for biaxiality degree $\Delta=4/5$ and in the absence of isotropic interaction ($U=0$). $\mathrm{ISO}$: isotropic phase. $\mathrm{N^{+}_{U}}$: calamitic uniaxial nematic phase. $\mathrm{N^{-}_{U}}$: discotic uniaxial nematic phase. $\mathrm{N_{B}}$: biaxial nematic phase. Short-dashed lines are the boundaries of biphasic regions (gray). Red dash-dotted line: critical end point ($\mathrm{CE}$). Short-dotted red line in the insets mark triple points. $\mathrm{TC}$ is a tricritical point.}
    \label{fig:434}
\end{figure}

For intrinsically biaxial nematogens with $\Delta=4/5$, we can find a great variety of phase diagrams as a function of the parameter $B$. For $(A,B)=(1,1)$, we can see from the phase diagram in Fig.\,\ref{fig:434a} that, at high concentration of nematogens, the sequence of phases observed, as the temperature decreases, is a nematogen-rich $\mathrm{ISO}$ phase, followed by a first-order weak $\mathrm{ISO}$-$\mathrm{N^{+}_{U}}$ phase transition, then by a large temperature range where an $\mathrm{N^{+}_{U}}$ phase is stable, limited below by a second-order $\mathrm{N^{+}_U}$-$\mathrm{N_B}$ phase transition, and finally a low-temperature $\mathrm{N_B}$ phase is stable. 
The conditions for the second-order $\mathrm{N^{+}_U}$-$\mathrm{N_B}$ transition are determined by Eq.\,(\ref{eq:mfe}) supplemented by $\,d^{2}\psi/\,d\eta^{2}=0$ evaluated at $(S,\eta,R,\zeta,\phi)=(S_{o},0,R_{o},0,\phi_{o})$. 
At low temperature, the system presents a dipole-rich biaxial nematic phase. In this diagram, we have a critical end point ($\mathrm{CE}$), in which 
the second-order $\mathrm{N^{+}_U}$-$\mathrm{N_B}$ transition line meets the coexistence curves between the biaxial and the isotropic phases (at lower temperatures) and between the uniaxial and the isotropic phases (at intermediate temperatures).

When the strength of the anisotropic interaction between dipolar NPs and nematogens is such that it favors dipolar axes perpendicular to the first principal nematogenic axes, we obtain phase diagrams similar to the one shown in Fig.\,\ref{fig:434b}. The tricritical ($\mathrm{TC}$) point shown in the diagram, separating regions of continuous and first-order $\mathrm{N^{+}_U}$-$\mathrm{N_B}$ transitions, can be determined by Eq.\,(\ref{eq:mfe}), supplemented by $\,d^{2}\psi/\,d\eta^{2}=\,d^{4}\psi/\,d\eta^{4}=0$ evaluated at $(S,\eta,R, \zeta,\phi)=(S_{\mathrm{TC}},0,R_{\mathrm{TC}}, 0, \phi_{\mathrm{TC}})$, where $S_{\mathrm{TC}}$, $R_{\mathrm{TC}}$ and $\phi_\mathrm{TC}$ are, respectively, the values of $S$, $R$ and $\phi$ at the $\mathrm{TC}$ point. In addition to the $\mathrm{TC}$ point, the diagram presents a triple point (short-dotted red line in the inset), which indicates an $\mathrm{ISO}$-$\mathrm{N^{+}_U}$-$\mathrm{N_B}$ phase coexistence. Determining the location of the triple point requires Eq.\,(\ref{eq:mfe}) evaluated at $(S,\eta,R,\zeta,\phi)=(0,0,0,0,\phi_{\mathrm{I}})$, at $(S,\eta,R,\zeta,\phi)=(S_{\mathrm{U}},0,R_{\mathrm{U}},0,\phi_{\mathrm{U}})$, and at $(S,\eta,R,\zeta,\phi)=(S_{\mathrm{B}},0,R_{\mathrm{B}},0,\phi_{\mathrm{B}})$, supplemented by $\psi(0,0,0,0,\phi_{\mathrm{I}})=\psi(S_{\mathrm{U}},0,R_{\mathrm{U}},0,\phi_{\mathrm{U}})=\psi(S_{\mathrm{B}},0,R_{\mathrm{B}},0,\phi_{\mathrm{B}})$. At temperatures lower than the triple-point temperature, both $\mathrm{ISO}$-$\mathrm{N^{+}_U}$ and $\mathrm{N^{+}_U}$-$\mathrm{N_B}$ coexistences become metastable with respect to the $\mathrm{ISO}$-$\mathrm{N_B}$ first-order transition. For a fixed temperature at the range $T_{\mathrm{TC}}\leqslant T\lessapprox 0.5$, the sequence of phases observed, as the concentration increases, is an $\mathrm{ISO}$ phase, followed by a small $\mathrm{ISO}$-$\mathrm{N^{+}_U}$ coexistence region, then by a pure $\mathrm{N_B}$ phase limited by dipole-rich and dipole-poor $\mathrm{N^{+}_U}$-$\mathrm{N_B}$ second-order transitions on the left and on the right, respectively, and finally a stable $\mathrm{N^{+}_{U}}$ phase. 

Upon decreasing the value of $B$, while $T_{\mathrm{TC}}$ increases at the $\phi$-$T$ phase diagrams, a small region begins to appear where a $\mathrm{N^{-}_U}$ phase is stable. The small region is delimited by a second-order transition line (to the biaxial phase) ending in two $\mathrm{CE}$ points with temperatures $T^{(1)}_{\mathrm{CE}}$ and $T^{(2)}_{\mathrm{CE}}$, with $T^{(1)}_{\mathrm{CE}} \leqslant T^{(2)}_{\mathrm{CE}}$. For $B\approx-1.1868$, the triple point and the $\mathrm{CE}$ point with temperature $T^{(2)}_{\mathrm{CE}}$ meet at a higher-order multicritical point, while for $-108/91<B\lessapprox-1.1868$ the phase diagrams are qualitatively similar to that shown in Fig.\,\ref{fig:434c}. This new topology presents a $\mathrm{N^{+}_U}$-$\mathrm{N^{-}_U}$ phase transition limited from below by an $\mathrm{ISO}$-$\mathrm{N^{+}_U}$-$\mathrm{N^{-}_U}$ triple point and from above by a $\mathrm{CE}$ point; see the inset in Fig.\,\ref{fig:434c}.

\begin{figure}[ht]
    \centering
    \begin{minipage}[b]{0.475\textwidth}
     \centering
     \includegraphics[scale=0.9]{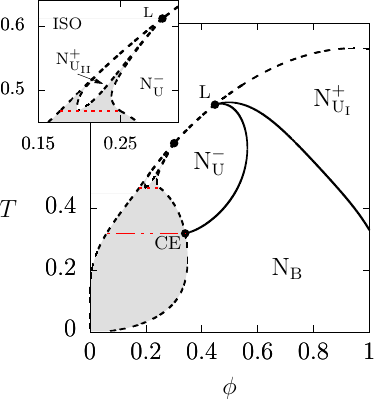}
     \subcaption{$(A, B)=(1,-6/5)$}
     \label{fig:4411a}
    \end{minipage}
    \hspace{10pt}
    \begin{minipage}[b]{0.475\textwidth}
     \centering
     \includegraphics[scale=0.9]{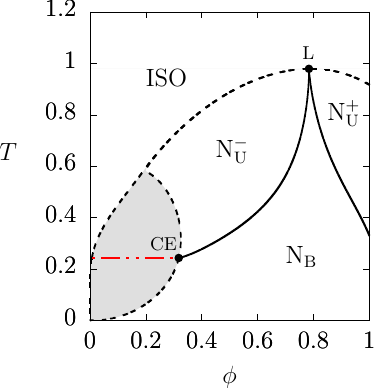}
     \subcaption{$(A, B)=(1,-3/2)$}
     \label{fig:4411b}
    \end{minipage}
    \hspace{10pt}
    \begin{minipage}[b]{0.475\textwidth}
     \centering
     \includegraphics[scale=0.9]{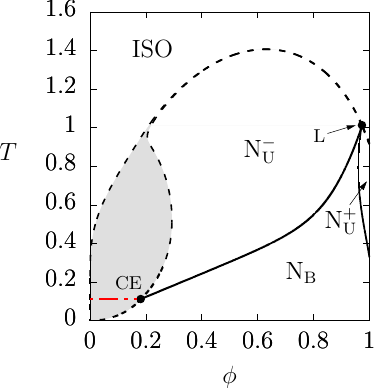}
     \subcaption{$(A, B)=(1,-27/10)$}
     \label{fig:4411c}
    \end{minipage}
    \caption{Phase diagrams in terms of temperature $T$ (in units of $A$) and concentration $\phi$ of nematogens, for biaxiality degree $\Delta=4/5$ and in the absence of isotropic interaction ($U=0$). $\mathrm{ISO}$: isotropic phase. $\mathrm{N^{+}_{U}}$, $\mathrm{N^{+}_{U_I}}$, and $\mathrm{N^{+}_{U_{II}}}$: calamitic uniaxial nematic phases. $\mathrm{N^{-}_{U}}$: discotic uniaxial nematic phase. $\mathrm{N_{B}}$: biaxial nematic phase. Short-dashed lines are the boundaries of biphasic regions (gray). Red dash-dotted line: critical end point ($\mathrm{CE}$). Short-dotted red line in the inset mark triple points}. $\mathrm{L}$ is a Landau multicritical point.
    \label{fig:4411}
\end{figure}

Landau points (discussed in Sec.~\ref{sec:lp}) are present in the $\phi$-$T$ phase diagrams depending of the strength of the anisotropic interaction between dipoles and nematogens. As we discussed for $0<\Delta<1$, we can find phase diagrams that can even present two Landau points. Fig.\,\ref{fig:4411a} represents the phase diagram for a system with $(A,B)=(1,-6/5)$, where we find different multicritical points and nematic phases. We can notice a wide region of stability for the $\mathrm{N_B}$ phase, bounded at high temperatures by $\mathrm{N^{-}_{U}}$-$\mathrm{N_B}$ and $\mathrm{N^{+}_{U_{I}}}$-$\mathrm{N_B}$ second-order lines. These two rather convoluted second-order-transition lines meet at a high-concentration $\mathrm{L}$ point,
at which the three nematic phases ($\mathrm{N_B}$, $\mathrm{N^+_{U_{I}}}$ and $\mathrm{N^{-}_U}$) become identical to each other and to the $\mathrm{ISO}$ phase. 
In the diagram we can see another $\mathrm{L}$ point, at low concentration, at which two uniaxial nematic phases ($\mathrm{N^{+}_{U_{II}}}$ and $\mathrm{N^{-}_U}$) become identical to the $\mathrm{ISO}$ phase. For temperatures below the low-concentration $\mathrm{L}$ point there are 
$\mathrm{ISO}$-$\mathrm{N^{+}_{U_{II}}}$ and $\mathrm{N^{+}_{U_{II}}}$-$\mathrm{N^{-}_U}$ biphasic regions.
The $\mathrm{N^{+}_{U_{II}}}$-$\mathrm{N^{-}_U}$ first-order transition can be determined by Eq.\,(\ref{eq:mfe}) evaluated at $(S, \eta,R,\zeta,\phi)=(S_{+},0,R_{+},0,\phi_{+})$ and at $(S, \eta,R,\zeta,\phi)=(S_{-},0,R_{-},0,\phi_{-})$, supplemented by $\psi(S_{+},0,R_{+},0,\phi_{+})=\psi(S_{-},0,R_{-},0,\phi_{-})$, where $\phi_{+}$ and $\phi_{-}$ are the nematogen concentrations at the $\mathrm{N^{+}_{U_{II}}}$ and $\mathrm{N^{-}_U}$ phases respectively, while $S_{\pm}$ and $R_{\pm}$ are the corresponding values of $S$ and $R$. Near the low-concentration $\mathrm{L}$ point, $\phi_{+}\to\phi_{-}$ and the uniaxial-uniaxial first-order transition is very weak; see the inset in Fig.\,\ref{fig:4411a}. There is an $\mathrm{ISO}$-$\mathrm{N^{+}_{U_{II}}}$-$\mathrm{N^{-}_U}$ triple point (red short-dotted line) which marks the limit of stability for both  $\mathrm{ISO}$-$\mathrm{N^{+}_{U_{II}}}$ and $\mathrm{N^{+}_{U_{II}}}$-$\mathrm{N^{-}_U}$ coexistence regions because, for temperatures below the triple point, these coexistences become metastable with respect to the  $\mathrm{ISO}$-$\mathrm{N^{-}_{U}}$ coexistence.

In general, for values of $B<-108/91$, the phase diagrams for $\Delta=4/5$ present a stable high-concentration $\mathrm{L}$ point. Nevertheless the low-concentration $\mathrm{L}$ point is only stable for $-1.3645\lessapprox B<-108/91$, because, for values of $B<-1.3645$, the low-concentration $\mathrm{L}$ point represents a local minimum of the Landau--de Gennes free-energy functional but becomes metastable with respect to the $\mathrm{ISO}$-$\mathrm{N^{-}_{U}}$ coexistence. The loss of this Landau point obviously implies the disappearance of the region of stability of the $\mathrm{N^{+}_{U_{II}}}$ phase, thus the $\phi$-$T$ phase diagram adopts a topology similar to that represented in Fig.\,\ref{fig:4411b}, for $(A,B)=(1,-3/2)$. As $B$ is further reduced, an increase in the region where the $\mathrm{N^{-}_U}$ phase is stable and a decrease in the stability region of the $\mathrm{N^{-}_U}$ phase can be noted; see Fig.\,\ref{fig:4411c}. For $B\ll -A$, the $\mathrm{N^{+}_U}$ phase is stable only very close to the line $\phi=1$.

\subsection{Phase diagrams for \texorpdfstring{$\Delta=1$}{}}
\begin{figure}[ht]
    \centering
    \begin{minipage}[b]{0.475\textwidth}
     \centering
     \includegraphics[scale=0.9]{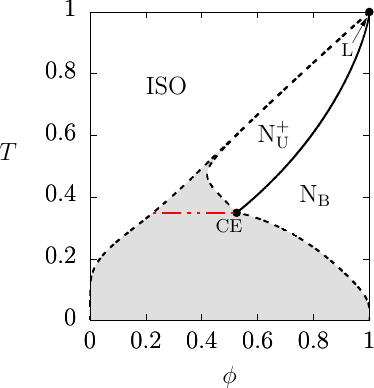}
     \subcaption{$(A, B)=(1,1/2)$}
     \label{fig:88a}
    \end{minipage}
    \hspace{10pt}
    \begin{minipage}[b]{0.475\textwidth}
     \centering
     \includegraphics[scale=0.9]{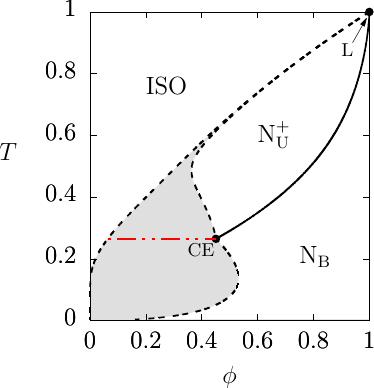}
     \subcaption{$(A, B)=(1,3/4)$}
     \label{fig:88b}
    \end{minipage}
    \hspace{10pt}
    \begin{minipage}[b]{0.475\textwidth}
     \centering
     \includegraphics[scale=0.9]{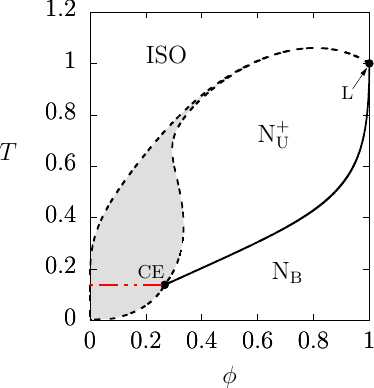}
     \subcaption{$(A, B)=(1,3/2)$}
     \label{fig:88c}
    \end{minipage}
    \caption{Phase diagrams in terms of temperature $T$ (in units of $A$) and concentration $\phi$ of nematogens, for biaxiality degree $\Delta=1$ and in the absence of isotropic interaction ($U=0$). $\mathrm{ISO}$: isotropic phase. $\mathrm{N^{+}_{U}}$: calamitic uniaxial nematic phase. $\mathrm{N_{B}}$: biaxial nematic phase. Short-dashed lines are the boundaries of biphasic region (gray). Red dash-dotted line: critical end point ($\mathrm{CE}$). $\mathrm{L}$ is a Landau multicritical point.}
    \label{fig:88}
\end{figure}

If the biaxiality degree of the nematogens is $\Delta=1$, we obtain phase diagrams such as those shown in Fig.\,\ref{fig:88}. The diagrams present a single multicritical $\mathrm{L}$ point located at $(\phi_{\mathrm{L}}, T_{\mathrm{L}})=(1,1)$, as discussed in Sec.~\ref{sec:lp}. These diagrams present a stable phase $\mathrm{N^{+}_{U}}$ whose stability region is strongly dependent on $B$. We know that for $B=0$ the system is characterized by the absence of the uniaxial phase, as, in this case, we have a diluted quadrupolar problem; see Ref.~\cite{oropesa2022phase}. For $B\gg A>0$, we find a large stability region for the $\mathrm{N^{+}_U}$ phase; see Fig.\,\ref{fig:88c}. In addition to the $\mathrm{L}$ point, these diagrams present a $\mathrm{CE}$ point that marks the stability limit of an $\mathrm{ISO}$-$\mathrm{N_B}$ coexistence region. As discussed in Appendix \ref{app:B}, for $T\to0$ there is a coexistence region between the $\mathrm{ISO}$ and the $\mathrm{N_B}$ phases only if $|B|<2/3$, and outside this range we find a stable $\mathrm{N_B}$ phase for all concentrations at zero temperature. Note that, for the sequence of diagrams from \ref{fig:88a} to \ref{fig:88c}, increasing $B$ leads to a reduction of the isotropic-nematic coexistence region.

Choosing $B<0$ with $\Delta=1$ leads to phase diagrams identical to the ones obtained for $B>0$, with the sole modification that calamitic phases change to discotic.

\section{LC-based suspensions}\label{sec:lcbs}For high concentration of nematogens ($\phi\to1$) the role of the dipolar NPs can be favorable or detrimental to the formation of nematic structures, as qualified by the increase or decrease of the clearing temperature. This is strongly dependent on the strength of the anisotropic interaction between dipolar NPs and nematogens. For uniaxial systems, the behavior of the doped transition temperature $T_{\mathrm{I}\text{-}\mathrm{U}}$ of the host liquid crystal with respect to the pure liquid-crystal transition temperature $T^{(0)}_{\mathrm{I}\text{-}\mathrm{U}}$ has been studied using different approaches. In the framework of the molecular MF theory, M. V. Gorkunov and M. A. Osipov~\cite{gorkunov2011mean} predict a softening of the isotropic-uniaxial first-order transition caused by strongly anisotropic interactions between nanoparticles and nematogens, together with a shift
$\delta T_{\mathrm{I}\text{-}\mathrm{U}}=T_{\mathrm{I}\text{-}\mathrm{U}}-T^{(0)}_{\mathrm{I}\text{-}\mathrm{U}}$ 
in the isotropic-uniaxial transition temperature
that can be positive or negative depending on the anisotropic interaction between objects of different nature and on the concentration of NPs. 
Similar results are obtained by A. N. Zakhlevnykh \textit{et al.}~\cite{zakhlevnykh2016simple} using the spherical approximation to derive an analytic expression for the free-energy functional.
Experimentally, negative and positive shifts in the transition temperature have been reported for various systems~\cite{selevou20178ocb, tomavsovivcova2015magnetically,  vimal2016thermal, orlandi2016doping, mishra2015electrical, kurochkin2010nano}. Using our LGMSZ-B model, we present below a detailed analysis of both uniaxial-isotropic and uniaxial-biaxial transitions in the limit of an LC-based suspension ($\phi\geqslant 0.9$).

\begin{figure}
    \centering
    \begin{minipage}[b]{0.475\textwidth}
     \centering
     \includegraphics[scale=0.9]{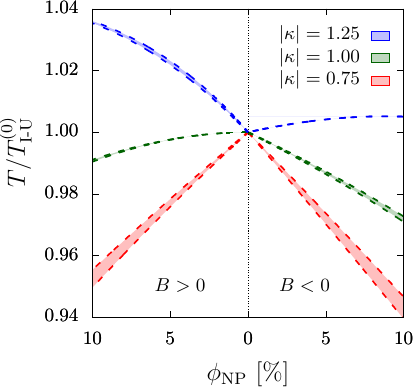}
     \subcaption{}
     \label{fig:85a}
    \end{minipage}
    \hspace{10pt}
    \begin{minipage}[b]{0.475\textwidth}
     \centering
     \includegraphics[scale=0.9]{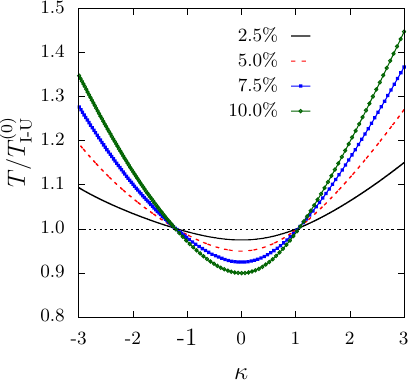}
     \subcaption{}
     \label{fig:85b}
    \end{minipage}
    \hspace{10pt}
    \caption{$(a)$ Coexistence regions in the plane of normalized isotropic-uniaxial transition temperature versus concentration $\phi_{\mathrm{NP}}$ of dipolar NPs for intrinsically uniaxial nematogens ($\Delta=0$) and different values of $\kappa=B/A$.  $(b)$ Normalized isotropic-uniaxial transition temperature as a function of $\kappa=B/A$ for intrinsically uniaxial nematogens ($\Delta=0$) and different values of $\phi_{\mathrm{NP}}$. Black dotted line indicates the clearing temperature ($T=T^{(0)}_{\mathrm{I}-\mathrm{U}}$) of the LC host.}
    \label{fig:85}
\end{figure}

 The simplest situation are the LC-based suspensions with rod-like ($\Delta=0$) nematogens, which can only present an $\mathrm{ISO}$-$\mathrm{N^{+}_U}$ first-order phase transition; see Sec.~\ref{sec:pdd0}. Fig.\,\ref{fig:85a} shows the $\mathrm{ISO}$-$\mathrm{N^{+}_U}$ coexistence region for different values of $\kappa=B/A$ and $\Delta=0$, with the temperature normalized by $T^{(0)}_{\mathrm{I}\text{-}\mathrm{U}}=\tfrac{9}{16\ln 2}$,
the transition temperature in the pure liquid-crystal system. As the concentration $\phi_\mathrm{NP}$ of nanoparticles is allowed to vary at a fixed temperature, phase separation occurs for a range of values of $\phi_\mathrm{NP}$, giving rise to the coexistence curves.
For a given value of $\phi_\mathrm{NP}$, we can see a reduction of the coexistence region as $|\kappa|$ increases. This reduction represents a softening of the phase transition.

The relative transition temperature as a function of $\kappa$ for different values of fixed concentration $\phi_{\mathrm{NP}}$ of dipolar NPs is shown in Fig.\,\ref{fig:85b}. For $\kappa=0$, we have a diluted problem~\cite{oropesa2022phase}, and in this case $\delta T_{\mathrm{I}\text{-}\mathrm{U}}<0$ because the presence of dipolar NPs decreases the effective interaction between nematogens. For $\kappa>0$, the dipoles tend to align with the first axes of the nematogens, that is, the macroscopic director $\hat{\bm{\mathrm{n}}}$ corresponds to an {easy axis} for the dipoles. This orientational preference increases the effective interaction between nematogens but implies a decrease in the dipole contribution to the entropy of the system. For $\kappa<0$, the dipoles tend to arrange themselves perpendicularly to $\hat{\bm{\mathrm{n}}}$, corresponding to an {easy plane} for the dipoles. Naturally, this orientational restriction is less costly, from an entropic point of view, than the one discussed for $\kappa>0$. The behavior of the system is qualitatively similar to the case $\kappa>0$, as we can see in Fig.\,\ref{fig:85b}.

For intrinsically biaxial nematogens (e.g., $\Delta=4/5$), in the limit of LC suspensions we will not only have an $\mathrm{ISO}$-$\mathrm{N^{+}_U}$ first-order transition, but also an $\mathrm{N^{+}_U}$-$\mathrm{N_B}$ (uniaxial-biaxial) second-order transition. In the case of pure LC, the transition temperatures are $T^{(0)}_{\mathrm{I}\text{-}\mathrm{U}}\approx0.9171$ and $T^{(0)}_{\mathrm{U}\text{-}\mathrm{B}}\approx0.3304$.  
\begin{figure}
    \centering
    \begin{minipage}[b]{0.475\textwidth}
     \centering
     \includegraphics[scale=0.9]{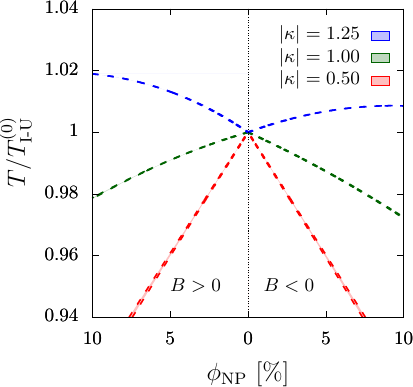}
     \subcaption{}
     \label{fig:464a}
    \end{minipage}
    \hspace{10pt}
    \begin{minipage}[b]{0.475\textwidth}
     \centering
     \includegraphics[scale=0.9]{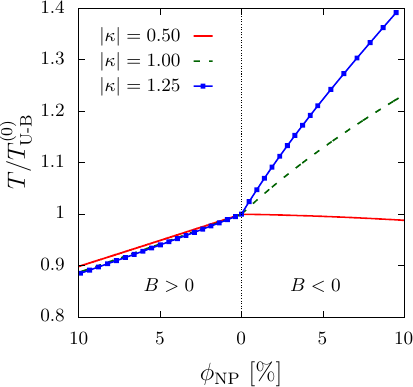}
     \subcaption{}
     \label{fig:464b}
    \end{minipage}
    \hspace{10pt}
    \caption{$(a)$ Normalized isotropic-uniaxial transition temperature as a function of the concentration of dipolar NPs $\phi_{\mathrm{NP}}$ for intrinsically biaxial nematogens $\Delta=4/5$ and different values of $\kappa=B/A$. $(b)$ Normalized uniaxial-biaxial transition temperature as a function of $\phi_{\mathrm{NP}}$ for intrinsically biaxial nematogens $\Delta=4/5$ and different values of $\kappa=B/A$.}
    \label{fig:464}
\end{figure}
Fig.\,\ref{fig:464} shows the temperatures of the two phase transitions for different values of $\kappa$. Similar to the case for rod-like nematogens, the $\mathrm{ISO}$-$\mathrm{N^{+}_U}$ biphasic region is reduced by increasing $|\kappa|$, implying a softening of the first-order transition. For $\kappa>0$, we can see that $\delta T_{\mathrm{U}\text{-}\mathrm{B}}=T_{\mathrm{U}\text{-}\mathrm{B}}-T^{(0)}_{\mathrm{U}\text{-}\mathrm{B}}<0$ for the entire range $0<\phi_{\mathrm{NP}}<0.1$, while the case $\kappa<0$ represents a favorable situation for the formation of biaxial structures for $\kappa\lesssim-0.5115$, since $\delta T_{\mathrm{U}\text{-}\mathrm{B}}>0$ (exemplified in  Fig.\,\ref{fig:464b} by the blue (upper) and dark-green (middle) solid lines for $B<0$), but this behavior, as we will discuss below, can be affected by the existence of a high-concentration $\mathrm{L}$ point.

In Sec.~\ref{sec:lp}, and more specifically in Sec.~\ref{sec:pdd08} for the case of $\Delta=4/5$, we discussed the existence of a stable $\mathrm{L}$ point for $\Delta\ne0$. An $\mathrm{L}$ point similar to that shown in Fig.\,\ref{fig:4411b} is associated with the existence of converging $\mathrm{N^{-}_U}$-$\mathrm{N_B}$ and $\mathrm{N^{+}_U}$-$\mathrm{N_B}$ second-order lines, as well as $\mathrm{ISO}$-$\mathrm{N^{-}_U}$ and $\mathrm{ISO}$-$\mathrm{N^{+}_U}$ first-order transitions. The concentration $\phi_{\mathrm{L}}$ at the $\mathrm{L}$ point increases as $B$ decreases (for $\Delta=4/5$ fixed), and, for sufficiently negative values of $B$, this $\mathrm{L}$ point can approach the point $(\phi,T)=(1,T^{(0)}_{\mathrm{I}\text{-}\mathrm{U}})$, as in Fig.\,\ref{fig:4411c}. 

\begin{figure}
    \centering
    \includegraphics{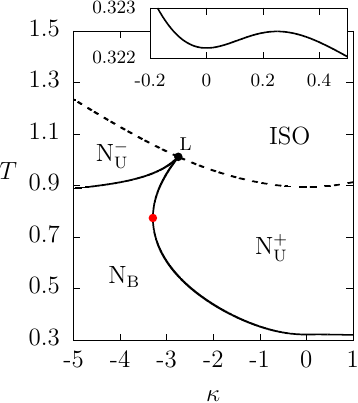}
    \caption{Phase diagram in terms of temperature $T$ (in units of $A$) and $\kappa=B/A$, for biaxiality degree $\Delta=4/5$, concentration of nematogens $\phi=0.975$, and in the absence of isotropic interaction ($U=0$). $\mathrm{ISO}$: isotropic phase. $\mathrm{N^{+}_{U}}$: calamitic uniaxial nematic phase. $\mathrm{N^{-}_{U}}$: discotic uniaxial nematic phase. $\mathrm{N_{B}}$: biaxial nematic phase. $\mathrm{L}$ is a Landau multicritical point. 
The red (light gray) point has coordinates $\left(\kappa_o, T_o\right)$.
}
    \label{fig:pdb}
\end{figure}

As noted throughout this Section, we define the LC-suspension limit as that in which nematogens represent at least $90\%$ of the mixture. If the value of $B$ implies that $\phi_{\mathrm{L}}\geqslant0.9$, then we will have a suspension that presents a stable $\mathrm{L}$ point. Fig.\,\ref{fig:pdb} shows the $\kappa$-$T$ phase diagram for an LC suspension with concentration of nanoparticles $\phi_{\mathrm{NP}}=0.025$ and nematogens with biaxiality degree $\Delta=4/5$. In the $\mathrm{N_B}$ phase, for $\kappa>0$, the probability of finding a dipole aligned with the first axes of the nematogens is very high, which increases the effective interaction between nematogens. This alignment implies a decrease in the dipole 
contribution to the entropy, but this entropic cost is not very important for small values of $\kappa$. As the value of $B$ is increased, eventually the increase in the effective interaction between nematogens no longer compensates the decrease in the entropy of the system and, as a consequence, the temperature of the $\mathrm{N^{+}_U}$-$\mathrm{N_B}$ second-order transition decreases; see the inset in Fig.\,\ref{fig:pdb}. Note that in the $\mathrm{N^{+}_U}$ phase the suspension exhibits a cylindrical symmetry, which is broken in the $\mathrm{N_B}$ phase, making entropic effects on the free energy more important.

For $\kappa<0$, the system presents a multicritical $\mathrm{L}$ point where four phases ($\mathrm{N^{-}_U}$, $\mathrm{N^{+}_U}$, $\mathrm{N_B}$, and $\mathrm{ISO}$) become identical. On the second-order line corresponding to the $\mathrm{N_B}$-$\mathrm{N^{+}_U}$ transition, there is a point $(\kappa_{o}, T_{o})$ which satisfies the condition $\,dT/\,d\kappa\to\infty$. This point implies a reentrance of the $\mathrm{N_B}$ phase. 
In the conditions of Fig.\,\ref{fig:pdb}, we have $\kappa_{o}\approx -3.4$.

\begin{figure}
    \centering
    \begin{minipage}[b]{0.475\textwidth}
     \centering
     \includegraphics[scale=1.1]{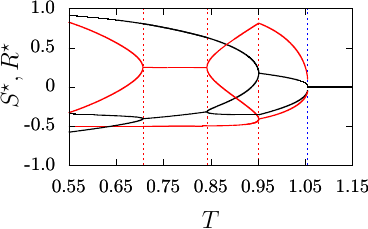}
     \subcaption{}
     \label{fig:ur88a}
    \end{minipage}
    \hspace{10pt}
    \begin{minipage}[b]{0.475\textwidth}
     \centering
     \includegraphics[scale=1.1]{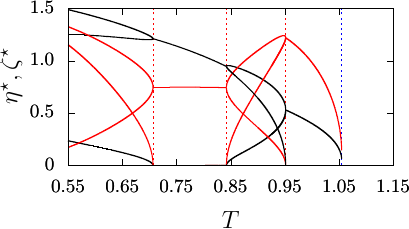}
     \subcaption{}
     \label{fig:ur88b}
    \end{minipage}
    \caption{Temperature dependence of the equilibrium values of (a) $S^{\star}=S/\phi$ (black solid line) and $R^{\star}=R/(1-\phi)$ (red or light gray solid line) and (b) $\eta^{\star}=\eta/\phi$ (black solid line) and $\zeta^{\star}=\zeta/(1-\phi)$  (red or light gray solid line), for $\kappa=-3.25$ and $\Delta=4/5$. Temperature scale is given in units of $A$. Red thin-dashed straight lines (three leftmost lines in each plot) signal second-order transitions and blue thin-dashed straight lines (rightmost in each plot) signal first-order transitions.}
    \label{fig:ur88}
\end{figure}

For fixed values of $\kappa$ such that $\kappa_{o}<\kappa<\kappa_{\mathrm{L}}$, the sequence of phases observed, as the temperature decreases, is $\mathrm{ISO}$, followed by an $\mathrm{N^{-}_{U}}$ phase where $S<0$ and $R>0$, followed by a small stability region for the $\mathrm{N_B}$ phase, bounded above and below by two second-order lines, followed by a $\mathrm{N^{+}_U}$ phase where $S>0$ and $R<0$, and finally the $\mathrm{N_B}$ is once more stable. This interesting phase behavior is also reflected in the scalar parameters $S$, $R$, $\eta$, and $\zeta$, as we can see in Fig.\,\ref{fig:ur88}. Notice that, except at zero temperature (not shown), biaxial solutions require that both $\eta$ and $\zeta$ take nonzero values. As the dipolar NPs are uniaxial, a nonzero $\zeta$ implies that there are nonzero fractions of dipoles pointing along each nematic director, and not only along the main one.

For values of $\kappa<\kappa_{o}$, as $T$ decreases, the phase sequence observed is $\mathrm{ISO}$, followed by a small stability region for the $\mathrm{N^{-}_U}$ phase, and finally a long stability region for the $\mathrm{N_B}$ phase. 

\section{Phase behavior for \texorpdfstring{$U\neq0$}{}}\label{sec:unoo}
For systems with nonzero isotropic lattice interaction ($U\ne0$), the phase diagrams can undergo various modifications. In the attractive regime, $U<0$, the regions of phase coexistence are enlarged. For sufficiently attractive values of $U$, we observe coexistence between dipole-rich and dipole-poor $\mathrm{ISO}$ phases. 

\begin{figure}
    \centering
    \begin{minipage}[b]{0.475\textwidth}
     \centering
     \includegraphics[scale=0.9]{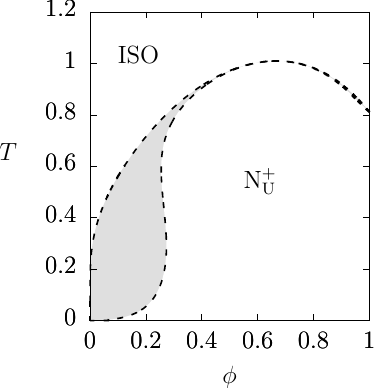}
     \subcaption{$(A, B, U)=(1,-2, -1)$}
     \label{fig:u88a}
    \end{minipage}
    \hspace{10pt}
    \begin{minipage}[b]{0.475\textwidth}
     \centering
     \includegraphics[scale=0.9]{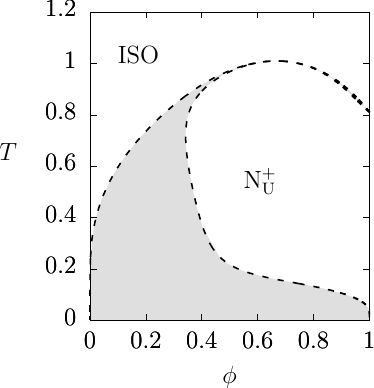}
     \subcaption{$(A, B, U)=(1,-2, -2)$}
     \label{fig:u88b}
    \end{minipage}
    \hspace{10pt}
    \begin{minipage}[b]{0.475\textwidth}
     \centering
     \includegraphics[scale=0.9]{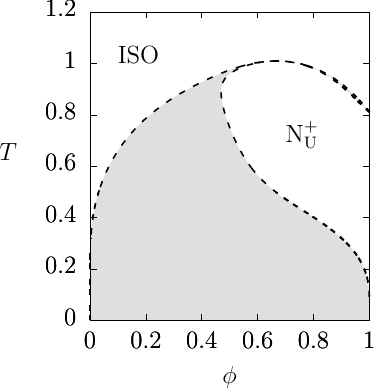}
     \subcaption{$(A, B, U)=(1,-2, -3)$}
     \label{fig:u88c}
    \end{minipage}
    \caption{Phase diagrams in terms of temperature $T$ (in units of $A$) and concentration $\phi$ of nematogens, for intrinsically uniaxial nematogens ($\Delta=0$) and different values of the isotropic interaction. $\mathrm{ISO}$: isotropic phase. $\mathrm{N^{+}_{U}}$: calamitic uniaxial nematic phase. Short-dashed lines are the boundaries of biphasic regions (gray).}
    \label{fig:u88}
\end{figure}

For $\Delta=0$, as $U$ is varied, we obtain the sequence of phase diagrams shown in Fig.\,\ref{fig:u88} for $(A,B)=(1,-2)$. Upon increasing the attractive character of $U$, we notice a reduction of the stability region of the $\mathrm{N^{+}_U}$ phase. At $T=0$, for $U<-3/2$, the $\mathrm{N^{+}_U}$ phase becomes metastable with respect to the $\mathrm{ISO}$-$\mathrm{N^{+}_U}$ coexistence. For a fixed temperature $T<1$, the width of the low-concentration $\mathrm{ISO}$-$\mathrm{N^{+}_U}$ coexistence region increases with $|U|$, as shown in Figs.~\ref{fig:u88a}-\ref{fig:u88c}.

\begin{figure}
    \centering
    \begin{minipage}[b]{0.475\textwidth}
     \centering
     \includegraphics[scale=0.9]{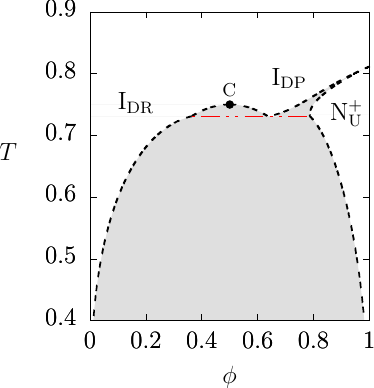}
     \subcaption{$(A, B, U)=(1,-1, -3)$}
     \label{fig:4e64a}
    \end{minipage}
    \hspace{10pt}
    \begin{minipage}[b]{0.475\textwidth}
     \centering
     \includegraphics[scale=0.9]{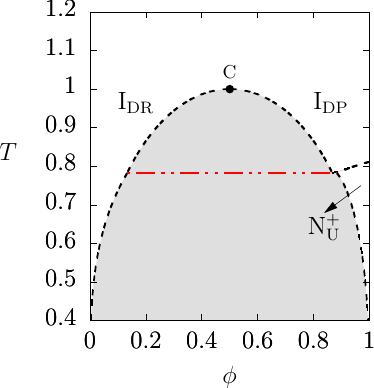}
     \subcaption{$(A, B, U)=(1,-1, -4)$}
     \label{fig:4e64b}
    \end{minipage}
    \hspace{10pt}
    \caption{Phase diagrams in terms of temperature $T$ (in units of $A$) and concentration $\phi$ of nematogens, for intrinsically uniaxial nematogens ($\Delta=0$) and for different values of the isotropic interaction. $\mathrm{I_{DR}}$: dipole-rich isotropic phase. $\mathrm{I_{DP}}$: dipole-poor isotropic phase. $\mathrm{N^{+}_{U}}$: calamitic uniaxial nematic phase. Short-dashed lines are the boundaries of biphasic regions (gray). Red dash-dotted line is an $\mathrm{I_{DR}}$-$\mathrm{I_{DP}}$-$\mathrm{N^{+}_U}$ triple point. $\mathrm{C}$ is a simple critical point.}
    \label{fig:4e64}
\end{figure}

For sufficiently attractive values of $U$, we can identify dipole-rich ($\mathrm{I_{DR}}$) and dipole-poor ($\mathrm{I_{DP}}$) isotropic phases. Fig.\,\ref{fig:4e64a} shows the $\phi$-$T$ phase diagram for a system with parameters $(A,B,U)=(1,-1,-3)$ and rod-like molecules. At low temperature, there is a coexistence region between a dipole-poor $\mathrm{N^{+}_U}$ phase and the $\mathrm{I_{DR}}$ phase. The diagram also exhibits an $\mathrm{I_{DR}}$-$\mathrm{I_{DP}}$-$\mathrm{N^{+}_U}$ triple point (red dash-dotted line), which marks the limit of stability of the $\mathrm{I_{DR}}$-$\mathrm{N^{+}_U}$ coexistence. The location of the triple point can be determined by the MF equations evaluated at $(S,\eta,R,\zeta, \phi)=(0,0,0,0,\phi_{\mathrm{I_{DR}}})$, at $(S,\eta,R,\zeta, \phi)=(0,0,0,0,\phi_{\mathrm{I_{DP}}})$, and at $(S,\eta,R,\zeta, \phi)=(S_{\mathrm{U}},0,R_{\mathrm{U}},0,\phi_{\mathrm{U}})$, supplemented by $\psi(0,0,0,0,\phi_{\mathrm{I_{DR}}})=\psi(0,0,0,0,\phi_{\mathrm{I_{DP}}})=\psi(S_{\mathrm{U}},0,R_{\mathrm{U}},0,\phi_{\mathrm{U}})$. There is a simple $\mathrm{C}$ point at which the $\mathrm{I_{DR}}$ and $\mathrm{I_{DP}}$ phases become identical. This simple $\mathrm{C}$ point can be determined by $\partial\psi/\partial\phi=\,d^{2}\psi/\,d\phi^{2}=\,d^{3}\psi/\,d\phi^{3}=0$ evaluated at $(S,\eta,R,\zeta,\phi)=(0,0,0,0,\phi_{\mathrm{C}})$, which leads to $\phi_{\mathrm{C}}=1/2$, $\beta_{\mathrm{C}}=-4/U$, and $\mu_{\mathrm{C}}=U/2$. 
The chemical potential remains equal to $\mu_{\mathrm{C}}$ for any point along the coexistence line between the $\mathrm{I_{DR}}$ and the $\mathrm{I_{DP}}$ phases.
In the context of the Landau--de Gennes free-energy functional, for attractive isotropic interactions, the simple $\mathrm{C}$ point can be either stable (an absolute minimum) or metastable (a local minimum). For repulsive isotropic interactions, $\mathrm{C}$ is unstable. 
If the isotropic interaction is more attractive (as in Fig.\,\ref{fig:4e64b}), there is an enlargement of the $\mathrm{I_{DR}}$-$\mathrm{I_{DP}}$ coexistence region and the simple $\mathrm{C}$ point takes a temperature greater than the $\mathrm{I_{DP}}$-$\mathrm{N^{+}_U}$ transition temperature for the pure LC host.

\begin{figure}
    \centering
    \begin{minipage}[b]{0.475\textwidth}
     \centering
     \includegraphics[scale=0.9]{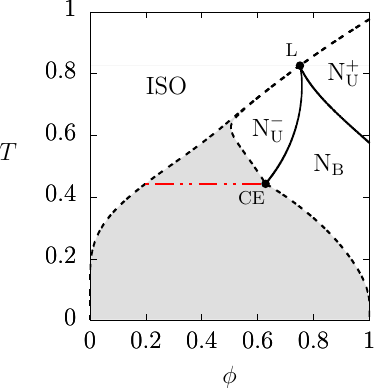}
     \subcaption{$(A, B, U)=(1,-3/4,-1)$}
     \label{fig:43e64a}
    \end{minipage}
    \hspace{10pt}
    \begin{minipage}[b]{0.475\textwidth}
     \centering
     \includegraphics[scale=0.9]{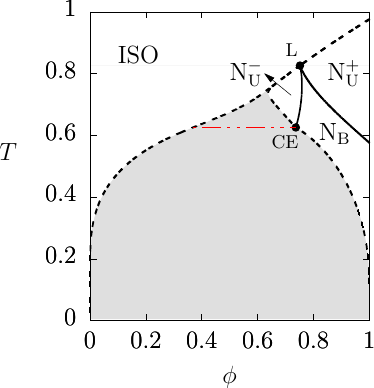}
     \subcaption{$(A, B, U)=(1,-3/4,-2)$}
     \label{fig:43e64b}
    \end{minipage}
    \hspace{10pt}
    \caption{Phase diagrams in terms of temperature $T$ (in units of $A$) and concentration $\phi$ of nematogens, for biaxiality degree $\Delta=19/20$ and different values of the isotropic interaction ($U=0$). $\mathrm{ISO}$: isotropic phase. $\mathrm{N^{+}_{U}}$: calamitic uniaxial nematic phase. $\mathrm{N^{-}_{U}}$: discotic uniaxial nematic phase. $\mathrm{N_{B}}$: biaxial nematic phase. Short-dashed lines are the boundaries of biphasic regions (gray). Red dash-dotted line: critical end point ($\mathrm{CE}$). $\mathrm{L}$ is a Landau point.}
    \label{fig:43e64}
\end{figure}

In the case of intrinsically biaxial nematogens with $\Delta=19/20$, the $\phi$-$T$ phase diagram for fixed parameters $(A,B,U)=(1,-3/4,-1)$ is shown in Fig.\,\ref{fig:43e64a}. This diagram presents a multicritical $\mathrm{L}$ point, located at $\phi_{\mathrm{L}}\approx0.7515$ and $T_{\mathrm{L}}\approx 0.8262$, which signals a direct transition from the $\mathrm{N_B}$ phase to the $\mathrm{ISO}$ phase. The position of the $\mathrm{L}$ point is independent of $U$. At high concentration of nematogens ($\phi>\phi_{\mathrm{L}}$), the phase sequence observed is $\mathrm{ISO}$, followed by a narrow $\mathrm{ISO}$-$\mathrm{N^{+}_U}$ coexistence, followed by a pure $\mathrm{N^{+}_U}$ phase that is bounded from below by a second-order line, followed by a pure $\mathrm{N_B}$ phase, and finally a low temperature coexistence between a dipole-poor $\mathrm{N_B}$ and a dipole-rich $\mathrm{ISO}$ phase. For values of $\phi_{\mathrm{CE}}<\phi<\phi_{\mathrm{L}}$, as $T$ decreases, the observed phase sequence is analogous, but the uniaxial phase is discotic. The $\phi$-$T$ phase diagram shown in Fig.\,\ref{fig:43e64b} illustrates a similar situation (with respect to Fig.\,\ref{fig:43e64a}), but in this case $U=-2$. We can observe that stronger attractive isotropic interactions reduce the stability regions of the $\mathrm{N_B}$ and the $\mathrm{N^{-}_U}$ phases. 

\begin{figure}
    \centering
    \begin{minipage}[b]{0.475\textwidth}
     \centering
     \includegraphics[scale=0.9]{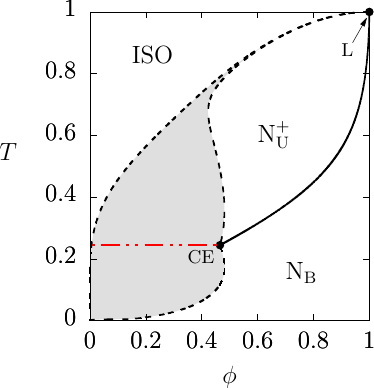}
     \subcaption{$(A, B, U)=(1,2/\sqrt{3}, -1)$}
     \label{fig:uee88a}
    \end{minipage}
    \hspace{10pt}
    \begin{minipage}[b]{0.475\textwidth}
     \centering
     \includegraphics[scale=0.9]{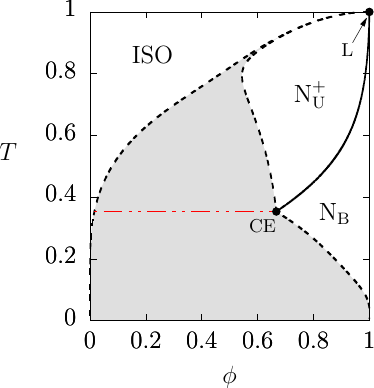}
     \subcaption{$(A, B, U)=(1,2/\sqrt{3}, -2)$}
     \label{fig:uee88b}
    \end{minipage}
    \hspace{10pt}
    \begin{minipage}[b]{0.475\textwidth}
     \centering
     \includegraphics[scale=0.9]{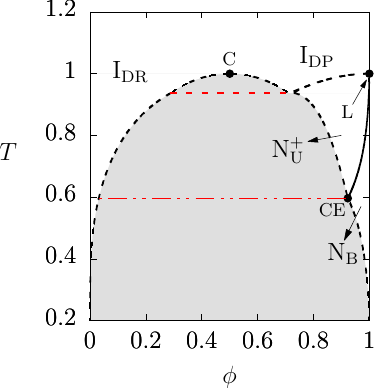}
     \subcaption{$(A, B, U)=(1,2/\sqrt{3}, -4)$}
     \label{fig:uee88c}
    \end{minipage}
    \caption{Phase diagrams in terms of temperature $T$ (in units of $A$) and concentration $\phi$ of nematogens, for biaxiality degree $\Delta=1$ and different values of the isotropic interaction. $\mathrm{ISO}$: isotropic phase. $\mathrm{I_{DR}}$: dipole-rich isotropic phase. $\mathrm{I_{DP}}$: dipole-poor isotropic phase.$\mathrm{N^{+}_{U}}$: calamitic uniaxial nematic phase. $\mathrm{N_{B}}$: biaxial nematic phase. Short-dashed lines are the boundaries of biphasic regions (gray). Red dash-dotted line: critical end point ($\mathrm{CE}$). Dotted red line marks a triple point. $\mathrm{L}$ is a Landau point. $\mathrm{C}$ is a simple critical point.}
    \label{fig:uee88}
\end{figure}

For maximally biaxial nematogens, $\Delta=1$, as we saw in Sec.~\ref{sec:lp}, the system presents a stable $\mathrm{L}$  point at $(\phi,T)=(1,1)$. In this case, it is impossible for two uniaxial phases to appear in the same phase diagram in the $\phi$-$T$ plane, regardless of the value of $U$. Fig.\,\ref{fig:uee88} shows a representative sequence of phase diagrams for a molecular system with parameters $(A,B)=(1,2/\sqrt{3})$ and different values for the strength of the isotropic interaction. The phase diagram in Fig.\,\ref{fig:uee88a} exhibits an $\mathrm{ISO}$-$\mathrm{N^{+}_U}$ coexistence region that widens as the temperature decreases. There is a $\mathrm{CE}$ point which is a stability bound because, for temperatures below $T_{\mathrm{CE}}$, the $\mathrm{ISO}$-$\mathrm{N^{+}_U}$ coexistence becomes metastable with respect to an $\mathrm{ISO}$-$\mathrm{N_B}$ coexistence. This diagram presents a second-order transition between an $\mathrm{N^{+}_U}$ and an $\mathrm{N_B}$ phase. The $\mathrm{N^{+}_U}$-$\mathrm{N_B}$ second-order line is limited at low and at high temperatures by $\mathrm{CE}$ and $\mathrm{L}$, respectively. The phase diagram shown in Fig.\,\ref{fig:uee88b} corresponds to a system with isotropic interaction $U=-2$. This diagram is similar to the one shown in Fig.\,\ref{fig:uee88a} with an important difference at low temperatures, namely, the $\mathrm{ISO}$-$\mathrm{N_B}$ coexistence region extends to $\phi=1$ as $T\to 0$. This is one more example of attractive isotropic interactions favoring isotropic-nematic phase coexistence. 

A sufficiently attractive isotropic interaction gives rise to the occurrence of a high-temperature $\mathrm{I_{DR}}$-$\mathrm{I_{DP}}$ coexistence ending at an ordinary $\mathrm{C}$ point, as shown Fig.\,\ref{fig:uee88c}. This diagram exhibits a $\mathrm{I_{DR}}$-$\mathrm{I_{DP}}$-$\mathrm{N^{+}_U}$ triple point. At temperatures below the triple-point temperature, the $\mathrm{I_{DR}}$-$\mathrm{I_{DP}}$ coexistence becomes metastable with respect to the $\mathrm{I_{DR}}$-$\mathrm{N^{+}_U}$ phase coexistence. There is also a $\mathrm{CE}$ point. The stability region for the $\mathrm{N_B}$ phase is very small because the isotropic interaction is much more important than both the anisotropic interaction between nematogens and the anisotropic interaction between dipolar NP and nematogens.

\section{Conclusions}\label{sec:conp}
We considered a mean-field model for a binary mixture of biaxial nematogens and dipolar nanoparticles, with Maier--Saupe-like interactions and discrete orientations, corresponding to the Zwanzig approximation. In addition to the anisotropic nematogen-nematogen and nematogen-nanoparticle interactions, we considered an isotropic interaction in the same spirit as the dilution problem discussed in Ref.~\cite{oropesa2022phase}. The Landau--de Gennes free-energy functional, the mean-field equations, and the position of many multicritical points (such as Landau points and simple critical points) were obtained exactly.

In the absence of isotropic interactions, we presented the concentration-temperature phase diagrams for fixed values of the biaxiality parameter $\Delta$. In this case, we found a large number of different phase-diagram topologies and various multicritical points, depending on the value of the nematogen-nanoparticle coupling, $B$, with the possible appearance of two stable Landau points. Other multicritical points, such as critical-end or tricritical points, are present; see e.g. Figs.~\ref{fig:434} and \ref{fig:4411}.

We presented a detailed investigation in the limit of liquid-crystal-based suspensions (i.e. systems with a low concentration of nanoparticles). For rod-like nematogens ($\Delta=0$), our results are capable of reproducing changes in the clearing temperature, for which there are experimental results; see e.g. Refs.~\cite{lin2015phase} and~\cite{kurochkin2010nano}. Our results are also compatible with mean-field calculations for models with continuous degrees of freedom, based on spherical approximations; see Ref.~\cite{khokhlov1985theory}. For intrinsically biaxial nematogens ($\Delta\ne0$), we studied the effects of the presence of nanoparticles on the uniaxial-biaxial second-order transition, showing that the corresponding temperature can either increase or decrease, depending on the value of $B$. We also showed that, again depending on the value of $B$, the isotropic-uniaxial first-order transition and the uniaxial-biaxial second-order transition lines may meet at a Landau multicritical point.

Systems with $U\ne0$ were analyzed and showed to exhibit a great variety of multicritical points, depending on the character of the isotropic interaction, the biaxiality degree, and the strength of the anisotropic interaction between objects of different nature. The main effect of attractive isotropic interactions is to enlarge the low-temperature coexistence region between the nematic and the isotropic phases.

Comparison between the full phase diagrams predicted by the present treatment and experimentally obtained phase diagrams would require determining the relation between our parameter $B$ and the concentrations of both nanoparticles and ion impurities in the solution; see e.g. Ref. \cite{Shoarinejad2021}. Future work should include the effect of coupling between nanoparticles. This is required to reproduce aspects such as the coexistence of magnetic and nematic order \cite{Mertelj2013,Liu2016}. 
As we neglected interactions between dipolar nanoparticles, our results for high concentration of those particles should be subject to further checks, and were presented here for the sake of completeness.

\begin{acknowledgments}
W. G. C. Oropesa and A. P. Vieira acknowledge financial support from the National Council for Scientific and Technological Development (CNPq – 465259/2014-6), the Coordination for the Improvement of Higher Education Personnel (CAPES), the National Institute of Science and Technology for Complex Fluids (INCT-FCx), and the S\~ao Paulo Research Foundation (FAPESP – 2014/50983-3 and FAPESP - 2023/04066-8). 
\end{acknowledgments}

\onecolumngrid

\appendix

\section{Mean-field calculations for the LGMSZ-B model}\label{app:A}
The MF version of the LGMSZ-B model is obtained by assuming a fully-connected lattice Hamiltonian
\begin{equation}
    \mathcal{H}_{\mathrm{mf}}=-\dfrac{A}{2N}\sum^{N}_{i,j=1}\gamma_{i}\gamma_{j}\bm{\Omega}_{i}\bm{:}\bm{\Omega}_{j}-\dfrac{B}{N}\sum^{N}_{i,j=1}\gamma_i(1-\gamma_j)\bm{\Omega}_{i}\bm{:}\bm{\Theta}_{j}+\dfrac{U}{2N}\sum_{i,j=1}^N\gamma_{i}\gamma_{j},
\end{equation}
where the sums over pairs of neighboring sites are replaced by sums over all pairs of sites, and the coupling parameters are replaced by new ones that are inversely proportional to the number of sites, to ensure that energy is extensive. The grand partition function is
\begin{equation}
    \Xi=\sum_{\{\gamma_{i}\}} \sum_{\{\bm{\Omega}_{i}\}} \sum_{\{\bm{\Theta}_{i}\}} \exp\left(\beta\mathcal{H}_{\mathrm{mf}}-\beta\mu\sum^{N}_{i=1}\gamma_{i}\right).
\end{equation}

In order to obtain an integral representation of the grand partition function in the MF limit, we introduce the two new variables
\begin{equation}
    \phi=\dfrac{1}{N}\sum^{N}_{i=1}\gamma_{i}\quad\text{and}\quad\bm{\mathrm{D}}=\dfrac{1}{N}\sum^{N}_{i=1}(1-\gamma_{i})\bm{\mathrm{\Theta}}_{i},
\end{equation}
with the help of Dirac delta functions, and use the Gaussian integral identity 
(analogous to the one used in the Hubbard--Stratonovich transformation \cite{Stratonovich1958,Hubbard1959}).
\begin{equation}
        \exp\left(\dfrac{\beta A}{2N}\sum_{i,j}\gamma_{i}\gamma_{j}\mathbf{\Omega}_{i}\mathbf{:}\mathbf{\Omega}_{j}\right)\propto\int\exp\left(-\dfrac{\beta AN}{2}\norm{\mathbf{Q}}^{2}
        +\beta A\sum_{i}\gamma_{i}\mathbf{Q}\mathbf{:}\mathbf{\Omega}_{i}\right)\,d[\mathbf{Q}],
        \label{eq:gaussian}
\end{equation}
in which $\mathbf{Q}=\mathrm{diag}\left(Q_{xx},Q_{yy},Q_{zz}\right)$  and $d[\mathbf{Q}]=dQ_{xx}dQ_{yy}dQ_{zz}$,
to obtain
\begin{equation}\label{eq:xin}
    \Xi\propto\int\limits_{\mathbb{R}^{7}}e^{-\beta N \Phi(\bm{\mathrm{Q}}, \bm{\mathrm{D}}, \phi)}\mathcal{D}(\bm{\mathrm{Q}}, \bm{\mathrm{D}}, \phi)\,d[\bm{\mathrm{Q}}]\,d[\bm{\mathrm{D}}]\,d\phi,
\end{equation}
with
\begin{equation}
    \Phi(\bm{\mathrm{Q}}, \bm{\mathrm{D}}, \phi)=
    \dfrac{A}{2}\norm{\bm{\mathrm{Q}}}^{2}+\dfrac{U}{2}\phi^{2}-\mu\phi,
\end{equation}
and
\begin{equation}\label{eq:de}
    \begin{split}
    \mathcal{D}(\bm{\mathrm{Q}}, \bm{\mathrm{D}}, \phi)&=\sum_{\{\bm{\Gamma}_{i}\}}\delta\left(\phi-\dfrac{1}{N}\sum^{N}_{i=1}\gamma_{i}\right)\delta\left(\bm{\mathrm{D}}-\dfrac{1}{N}\sum^{N}_{i=1}(1-\gamma_i)\right)\\
    &\quad\times\exp\left[-\beta(A\bm{\mathrm{Q}}+B\bm{\mathrm{D}})\bm{:}\sum^{N}_{i=1}\gamma_{i}\bm{\Omega}_{i}\right],
    \end{split}
\end{equation}
where the constant of proportionality in Eq.\,(\ref{eq:xin}) is irrelevant. 

Performing the sum in the exponential in Eq.\,(\ref{eq:de}) and using integral representations for the Dirac delta functions, we obtain
\begin{equation}\label{eq:F}
    \mathcal{D}(\bm{\mathrm{Q}}, \bm{\mathrm{D}}, \phi)=\left(\dfrac{N}{2\pi\im}\right)^{4}\int\limits_{\mathcal{C}}e^{-N F(\hat{\bm{\mathrm{D}}}, \hat{\phi})}\,d[\hat{\bm{\mathrm{D}}}]\,d\hat{\phi},
\end{equation}
where
\begin{equation}
     F(\hat{\bm{\mathrm{D}}}, \hat{\phi})=\hat{\bm{\mathrm{D}}}\bm{:}\bm{\mathrm{D}}+\hat{\phi}\phi-\ln\left[\chi(\hat{\bm{\mathrm{D}}}, \hat{\phi})\right],
\end{equation}
and
\begin{equation}
    \chi(\hat{\bm{\mathrm{D}}}, \hat{\phi})=e^{\hat{\phi}}\sum_{\{\bm{\Omega}\}}e^{\beta(A\bm{\mathrm{Q}}+B\bm{\mathrm{D}})\bm{:}\bm{\Omega}}+\sum_{\{\bm{\Theta}\}}e^{\hat{\bm{\mathrm{D}}}\bm{:}\bm{\Theta}}.
\end{equation}
In the thermodynamic limit ($N\gg1$), we can use the multivariable Morse lemma to get an extension of the steepest-descent method into complex variables $(\hat{\bm{\mathrm{D}}}, \hat{\phi})$; see Refs.~\cite{wong2001asymptotic} and~\cite{fedoryuk1989asymptotic}. We expect the integral in Eq.\,(\ref{eq:F}) to be dominated by the highest saddle point, $(\hat{\bm{\mathrm{D}}}_{o}, \hat{\phi}_{o})\in\mathcal{M}_{3}(\mathbb{C})\times\mathbb{C}$ \footnote{$\mathcal{M}_{n}(\mathbb{C})$ is the set of square matrices of order $n$ whose elements belong to the set $\mathbb{C}$.}, which can be determined from the conditions
\begin{equation}\label{eq:spa1}
    \dfrac{\partial}{\partial\hat{\bm{\mathrm{D}}}}F(\hat{\bm{\mathrm{D}}}_{o}, \hat{\phi}_{o})=0\quad\Rightarrow\quad\bm{\mathrm{D}}=\dfrac{1}{\chi(\hat{\bm{\mathrm{D}}}_{o}, \hat{\phi}_{o})}\sum_{\{\bm{\Theta}\}}e^{\hat{\bm{\mathrm{D}}}_{o}\bm{:}\bm{\Theta}}\bm{\Theta},
\end{equation}
and
\begin{equation}\label{eq:spa2}
    \dfrac{\partial}{\partial\hat{\phi}}F(\hat{\bm{\mathrm{D}}}_{o}, \hat{\phi}_{o})=0\quad\Rightarrow\quad\phi=\dfrac{e^{\hat{\phi}_{o}}}{\chi(\hat{\bm{\mathrm{D}}}_{o}, \hat{\phi_{o}})}\sum_{\{\bm{\Omega}\}}e^{\beta(A\bm{\mathrm{Q}}+B\bm{\mathrm{D}})\bm{:}\bm{\Omega}}.
\end{equation}
Under regularity conditions (single non-degenerate saddle point) we have
\begin{equation}\label{eq:sd}
    \mathcal{D}(\bm{\mathrm{Q}}, \bm{\mathrm{D}}, \phi)=\left(\dfrac{2\pi}{N}\right)^{2}\dfrac{e^{-NF(\hat{\bm{\mathrm{D}}}_{o}, \hat{\phi}_{o})}}{\sqrt{\det\bm{\mathrm{H}}(\hat{\bm{\mathrm{D}}}_{o}, \hat{\phi}_{o})}}+\mathcal{O}(1/N),
\end{equation}
where $\det\bm{\mathrm{H}}(\hat{\bm{\mathrm{D}}}_{o}, \hat{\phi}_{o})$ is the Hessian matrix having eigenvalues $\{\lambda_{i}\}$ with $i=\overline{1,4}$ and $|\arg(\lambda_{i})|<\pi/2$~\footnote{Note that \begin{equation*}
    \sqrt{\det\bm{\mathrm{H}}(\hat{\bm{\mathrm{D}}}_{o}, \hat{\phi}_{o})}=e^{\im\varphi/2}\prod^{4}_{k=1}|\lambda_{k}|^{1/2}\neq0,\quad\text{where}\quad \varphi=\sum_{k=1}^{4}\arg(\lambda_{k}).
\end{equation*}
If $(\hat{\bm{\mathrm{D}}}, \hat{\phi})\in\mathcal{M}_{3}(\mathbb{R})\times\mathbb{R}$ and $\Im[F(\bm{\mathrm{Q}}, \bm{\mathrm{D}}, \phi)]=0$ then $\varphi=0$. Also if  $(\hat{\bm{\mathrm{D}}}, \hat{\phi})\in\mathcal{M}_{3}(\mathbb{R})\times\mathbb{R}$, $\Re[F(\bm{\mathrm{Q}}, \bm{\mathrm{D}}, \phi)]=0$ and $|\sqrt{\lambda_{k}}|<\pi/4$ then $\varphi=\pi m/2$ where $m$ is the number of negative eigenvalues minus the number of positives ones (stationary-phase method).}. 

Now the grand partition function can be written as
\begin{equation}
    \Xi\propto\int\limits_{\mathbb{R}^{7}}e^{-\beta N\psi(\bm{\mathrm{Q}}, \bm{\mathrm{D}}, \phi)}\,d[\bm{\mathrm{Q}}]\,d[\bm{\mathrm{D}}]\,d\phi,
    \label{eq:Xi}
\end{equation}
where 
\begin{equation}
\psi(\bm{\mathrm{Q}}, \bm{\mathrm{D}}, \phi)=\Phi(\bm{\mathrm{Q}}, \bm{\mathrm{D}}, \phi) + \dfrac{1}{\beta}F(\hat{\bm{\mathrm{D}}}_{o}, \hat{\phi}_{o})
\label{eq:LdG}
\end{equation} 
is the Landau--de Gennes free-energy functional. Note that
Eqs.\,(\ref{eq:spa1}) and (\ref{eq:spa2}) implicitly yield $\hat{\bm{\mathrm{D}}}_{o}$ and $\hat{\phi}_{o}$ as functions of $\bm{\mathrm{Q}}$, $\bm{\mathrm{D}}$ and $\phi$.
As discussed, for instance, in Refs.\,\cite{Katriel1986,Ball2010}, $\psi(\bm{\mathrm{Q}}, \bm{\mathrm{D}}, \phi)$ could be used to provide a \emph{bona-fide} Landau--de Gennes expansion for our model.

The steepest-descent method can be applied once more, now to Eq.\,(\ref{eq:Xi}), yielding the MF equations for the equilibrium values of $\bm{\mathrm{Q}}$, $\bm{\mathrm{D}}$ and $\phi$ given the interaction strengths $A$, $B$ and $U$ and the parameters $T$ and $\mu$. These equations are
\begin{equation}
    \dfrac{\partial}{\partial\bm{\mathrm{Q}}}\psi(\bm{\mathrm{Q}},\bm{\mathrm{D}}, \phi)=0\quad\Rightarrow\quad \bm{\mathrm{Q}}=\dfrac{e^{\hat{\phi}_{o}}}{\chi(\hat{\bm{\mathrm{D}}}_{o}, \hat{\phi_{o}})}\sum_{\{\bm{\Omega}\}}e^{\beta(A\bm{\mathrm{Q}}+B\bm{\mathrm{D}})\bm{:}\bm{\Omega}}\bm{\Omega},
    \label{eq:mf1}
\end{equation}
\begin{equation}
    \dfrac{\partial}{\partial\bm{\mathrm{D}}}\psi(\bm{\mathrm{Q}},\bm{\mathrm{D}}, \phi)=0\quad\Rightarrow\quad \hat{\bm{\mathrm{D}}}_{o}=\dfrac{Be^{\hat{\phi}_{o}}}{\chi(\hat{\bm{\mathrm{D}}}_{o}, \hat{\phi_{o}})}\sum_{\{\bm{\Omega}\}}e^{\beta(A\bm{\mathrm{Q}}+B\bm{\mathrm{D}})\bm{:}\bm{\Omega}}\bm{\Omega},
    \label{eq:mf2}
\end{equation}
and
\begin{equation}
    \dfrac{\partial}{\partial\phi}\psi(\bm{\mathrm{Q}},\bm{\mathrm{D}}, \phi)=0\quad\Rightarrow\quad\hat{\phi}_{o}=\dfrac{e^{\hat{\phi}_{o}}}{\chi(\hat{\bm{\mathrm{D}}}_{o}, \hat{\phi_{o}})}\sum_{\{\bm{\Omega}\}}e^{\beta(A\bm{\mathrm{Q}}+B\bm{\mathrm{D}})\bm{:}\bm{\Omega}}.
    \label{eq:mf3}
\end{equation}

From Eqs.\,(\ref{eq:gaussian}) and (\ref{eq:mf1}), we conclude that the equilibrium value of $\bm{\mathrm{Q}}$ is associated with the biaxial-nematogen order parameter, and that it is a traceless tensor, which can be parametrized as
\begin{equation}
\bm{\mathrm{Q}} = 
\frac{1}{2}
    \begin{pmatrix}
    -S - \eta & 0 & 0 \\
    0 & -S + \eta & 0 \\
    0 & 0 &  2S
    \end{pmatrix}.
\end{equation}
As
Eq.\,(\ref{eq:spa1}) implies $\Tr\bm{\mathrm{D}}=0$, we can also parameterize the second-rank tensor associated with the dipolar nanoparticles as
\begin{equation}
    \bm{\mathrm{D}}=\dfrac{1}{2}\begin{pmatrix}
    -R-\zeta & 0 & 0 \\
    0 & -R +\zeta & 0 \\
    0 & 0 & 2R
       \end{pmatrix}.
\end{equation}

Likewise, Eq.\,(\ref{eq:mf2}) implies that $\bm{\mathrm{D}}_{o}$ is a traceless diagonal tensor. Parametrizing it in the same fashion as $\bm{\mathrm{D}}$, with parameters $\hat{R}_{o}$ and $\hat{\zeta}_{o}$, and following a lengthy but straightforward algebraic calculation, we can solve Eqs.\,(\ref{eq:spa1}) and (\ref{eq:spa2}) explicitly for $\hat{R}_{o}$, $\hat{\zeta}_{o}$, and $\hat{\phi}_{o}$ as functions of $S$, $\eta$, $R$, $\zeta$, and $\phi$, allowing us to write the Landau--de Gennes free-energy functional as
\begin{equation}
\begin{split}
    \psi(S,R,\eta,\zeta,\phi)&=\dfrac{A}{4}(3S^{2}+\eta^{2})+\dfrac{U}{2}\phi^{2}-\mu\phi+\dfrac{\zeta-R}{3\beta}\ln\left[(R-1+\phi)^{2}-\zeta^{2}\right]\\
    &\qquad + \dfrac{1-\phi}{3\beta}\ln\left\{(1+2R-\phi)\left[(R-1+\phi)^{2}-\zeta^{2}\right]\right\}\\
    &\qquad +\dfrac{2R}{3\beta}\ln\left(1+2R-\phi\right)-\dfrac{2\zeta}{3\beta}\ln\left(1-R-\zeta-\phi\right)\\
    &\qquad -\dfrac{\phi}{\beta}\ln\left[\Lambda(S,R,\eta,\zeta)\right]+\dfrac{\phi}{\beta}\ln(6\phi)-\dfrac{2}{\beta}\ln(6),
\end{split}
\end{equation}
where
\begin{equation}
  \begin{split}
\Lambda(S, R, \eta, \zeta) &= 2\exp\left\{ -\frac{3\beta}{4}\left[A(S+\eta)+B(R+\zeta) \right]\right\} \\
&\quad\times\cosh{\left\{\dfrac{3\beta}{4}\left[A\left(S-\dfrac{\eta}{3}\right)+B\left(R-\dfrac{\zeta}{3}\right)\right]\Delta\right\}} \\
&\quad+ 2\exp\left\{ -\frac{3\beta}{4}\left[A(S-\eta)+B(R-\zeta) \right]\right\} \\
&\quad\times\cosh{\left\{\dfrac{3\beta}{4}\left[A\left(S+\dfrac{\eta}{3}\right)+B\left(R+\dfrac{\zeta}{3}\right)\right]\Delta\right\}} \\
	&\quad + 2\exp\left[\frac{3\beta}{2}(AS+BR)\right] \cosh{\left[\dfrac{\beta}{2}(A\eta+B\zeta) \Delta\right]}.
\end{split}
\end{equation}
Minimizing $\psi(S,R,\eta,\zeta,\phi)$ with respect to its arguments yields manifestly self-consistent mean-field equations.

\section{The Landau point}\label{sec:lp}
Systems with quadrupole interactions may exhibit Landau  $(\mathrm{L})$ multicritical points (see, e.g., Ref~\cite{nascimento2015maier}), at which the isotropic phase and various nematic phases become identical. In order to determine the location of a Landau point, we must impose certain conditions on the Landau--de Gennes free-energy functional, that is, we need to satisfy Eqs.\,(\ref{eq:mfe}) and the extra conditions $\,d^{2}\psi/\,dS^{2}=\,d^{3}\psi/\,dS^{3}=0$ evaluated at $(S,\eta,R,\zeta,\phi)=(0,0,0,0,\phi_{\mathrm{L}})$. For intrinsically biaxial nematogens, these conditions lead to
\begin{equation}\label{eq:lm}
    \phi_{\mathrm{L}} e^{-\beta(\mu-U\phi_{\mathrm{L}})}=1-\phi_{\mathrm{L}},
\end{equation}
\begin{equation}\label{eq:lce}
   -16+\beta(\Delta^{2}+3)\left[4A+3B^{2}\beta(1-\phi_{\mathrm{L}})\right]\phi_{\mathrm{L}}=0,
\end{equation}
and
\begin{equation}\label{eq:lc}
   64(\Delta^{2}-1)-B^{3}\beta^{3}(\Delta^{2}+3)^{3}\phi_{\mathrm{L}}^{2}(1-\phi_{\mathrm{L}})=0,
\end{equation}

Note that, for $B=0$, Eqs.\,(\ref{eq:lce}) and (\ref{eq:lc}) imply that the Landau point can only exist if $\Delta=1$, and the $\phi$-$T$ phase diagrams present a line of Landau points $A\beta\phi_{\mathrm{L}}=1$; see Refs.~\cite{rodrigues2020magnetic} and ~\cite{oropesa2022phase}. For $B\neq0$ we have many candidates for multicritical Landau points, satisfying
\begin{equation}\label{eq:lt}
    A\beta_\pm=\xi_{1}(\Delta,\kappa)\pm\xi_{2}(\Delta,\kappa),\quad\text{with}\quad\kappa=\dfrac{B}{A},
\end{equation}
functions $\xi_{1,2}(\Delta,\kappa)$ being
\begin{equation}
    \xi_{1}(\Delta,\kappa)=\dfrac{2}{9}\left[\dfrac{9}{3+\Delta^{2}}-\dfrac{6}{\kappa^{2}}+\dfrac{3+\Delta^{2}}{(\Delta^{2}-1)\kappa}\right],
\end{equation}
and
\begin{equation}
    \xi_{2}(\Delta,\kappa)=\sqrt{(\Delta^{2}+3)-\dfrac{12}{\kappa}(\Delta^{2}-1)}\dfrac{\left|\dfrac{(\Delta^{2}+3)^{2}}{\kappa}-9(\Delta^{2}-1)\right|}{(\Delta^{2}-1)(\Delta^{2}+3)^{3/2}}.
\end{equation}
We refrain from writing the lengthy expressions for the concentration values $\phi^{(1)}_{\mathrm{L}}$ and $\phi^{(2)}_{\mathrm{L}}$ associated with the inverse temperatures $A\beta_+$ and $A\beta_{-}$. Note that the solutions in Eq.\,(\ref{eq:lt}) are invalid for $\Delta=1$ because $\xi_{1,2}(\Delta,\kappa)\to\infty$. The cases  $12(\Delta^{2}-1)/(\Delta^{2}+3)<\kappa<0$ (with $0<\Delta<1$) and $0<\kappa<12(\Delta^{2}-1)/(\Delta^{2}+3)$ (with $1<\Delta<3$) represent unphysical situations, as they lead to values of $\beta_\pm$ having nonzero imaginary part. From Eqs.\,(\ref{eq:lm}) to (\ref{eq:lc}) we can see that, for $\Delta=1$, the location of the Landau point is defined by $\phi_{\mathrm{L}}=1$, $A\beta=1$, and $\mu\to\infty$.

\begin{figure}
    \centering
    \begin{minipage}[b]{0.48\textwidth}
     \centering
     \includegraphics[scale=1]{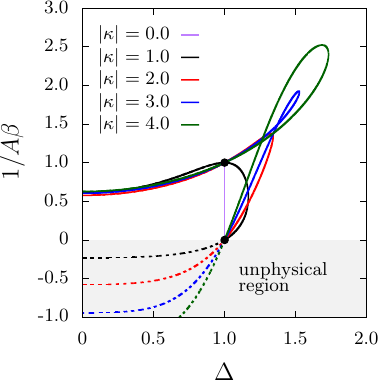}
     \hspace{20pt}
     \includegraphics[scale=1]{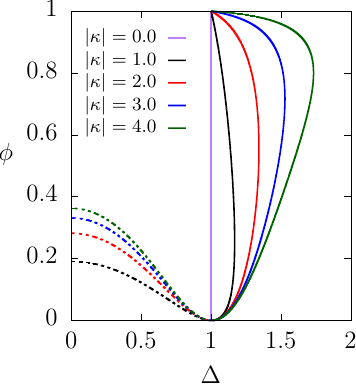}
     \subcaption{$B>0$}
     \label{fig:2132a}
    \end{minipage}
    \begin{minipage}[b]{0.48\textwidth}
     \centering
     \includegraphics[scale=1]{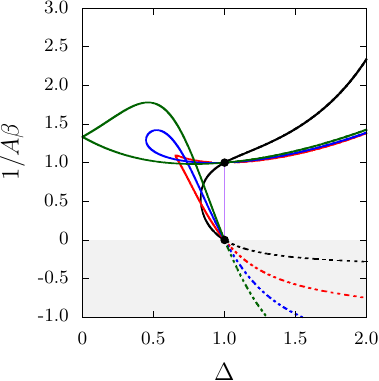}
     \hspace{20pt}     
     \includegraphics[scale=1]{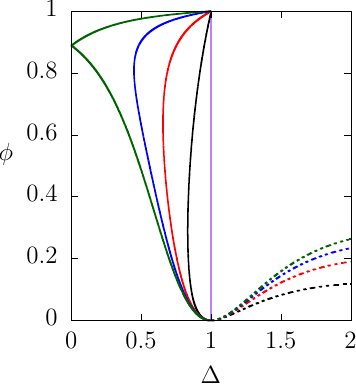}
     \subcaption{$B<0$}
     \label{fig:2132b}
    \end{minipage}
    \caption{
    Candidate multicritical Landau points in diagrams of temperature or nematogen concentration $\phi$ as functions of the molecular degree of biaxiality $\Delta$, for different values of $\kappa=B/A$. Gray region is unphysical. }
    \label{fig:2132}
\end{figure}

We emphasize that the solutions in Eq.\,(\ref{eq:lt}) represent candidate Landau points, whose stability still needs to be checked. However, those solutions indicate where we will not find Landau points. For $B>0$, see Fig.\,\ref{fig:2132a}, we can expect the following characteristics in the $\phi$-$T$ phase diagrams: $(i)$ for ``calamitic'' nematogens, $0<\Delta<1$, we find that $\xi_{1}(\Delta,\kappa)<\xi_{2}(\Delta,\kappa)$, leading to the unphysical solution $A\beta_{-} < 0$, so that the $\phi$-$T$ phase diagrams for these values of $\Delta$ can present at most one Landau point; $(ii)$ for ``discotic'' nematogens, $1<\Delta<3$, both solutions $A\beta_+$ and $A\beta_{-}$ are positive and in this case the $\phi$-$T$ phase diagrams can present one or two Landau multicritical points. 

On the other hand, if $B<0$, see Fig.\,\ref{fig:2132b}, for calamitic nematogens we can find $\phi$-$T$ phase diagrams with one or two Landau multicritical points. Finally, for discotic nematogens we have $\xi_{1}(\Delta,\kappa)+\xi_{2}(\Delta,\kappa)<0$, so that $A\beta_+<0$, and the $\phi$-$T$ phase diagrams can present at most one Landau point.

Irrespective of the value of $B$, the lower plots in Figs.\,\ref{fig:2132a} and \ref{fig:2132b} indicate that, for a fixed nematogen concentration $\phi$, a $T$-$\Delta$ phase diagram can present at most one Landau point.

In the particular case where $\xi_{2}(\Delta,\kappa)=0$ we have that $\beta_+=\beta_{-}$. The values of $\kappa$ for which this occurs are
\begin{equation}
    \kappa_{1}=\dfrac{12(\Delta^{2}-1)}{\Delta^{2}+3}\quad\Rightarrow\quad 
A\beta_{1}=\dfrac{(\Delta^{2}+3)^{3}+216(\Delta^{2}-1)^{2}}{108(\Delta^{2}-1)^{2}(\Delta^{2}+3)}\geqslant0
\end{equation}
and
\begin{equation}
    \kappa_{2}=\dfrac{(\Delta^{2}+3)^{2}}{9(\Delta^{2}-1)}\quad\Rightarrow\quad A\beta_{2}=\dfrac{4\Delta^{2}(\Delta^{2}-9)^{2}}{(\Delta^{2}+3)^{4}}>0.
\end{equation}

In order to analyze the stability of a Landau point, we must examine the higher order total derivatives of $\psi$ with respect to $\eta$, evaluated at $(S,\eta,R,\zeta,\phi)=(0,0,0,0,\phi_{\mathrm{L}})$. At the Landau point we will always have $\,d^{2}\psi/\,d\eta^{2}=0$, so that we need to look at the fourth-order derivative
\begin{equation}\label{eq:fod}
    \dfrac{\,d^{4}\psi}{\,d\eta^{4}}= \varsigma(\beta,\phi_{\mathrm{L}})-\dfrac{3(1-\phi_{\mathrm{L}})\left[16-3B^{2}\beta^{2}(\Delta^{2}+3)\phi_{\mathrm{L}}^{2}\right]^{2}}{64\beta(\Delta^{2}+3)^{2}\phi_{\mathrm{L}}^{3}\left[1+U\beta(1-\phi_{\mathrm{L}})\phi_{\mathrm{L}}\right]},
\end{equation}
with
\begin{equation}
     \varsigma(\beta,\phi_{\mathrm{L}})=-3\left[\dfrac{9B(\Delta^{2}-1)+A(\Delta^{2}+3)^{2}}{(\Delta^{2}+3)^{3}\phi^{2}_{\mathrm{L}}}-6\dfrac{18(\Delta^{2}-1)+(\Delta^{2}+3)^{3}}{\beta(\Delta^{2}+3)^{5}\phi^{3}_{\mathrm{L}}}\right].
\end{equation}
If the fourth-order derivative in Eq.\,(\ref{eq:fod}) is negative,
then the point does not represent a minimum of the Landau--de Gennes free-energy functional, therefore being unstable. (If the derivative is non-negative, the candidate point represents a minimum, but we must still check whether that minimum is absolute, before we can conclude that it represents a physically realizable Landau point.)

\section{Low-temperature coexistence}\label{app:B}
For this model, the free-energy of the system strongly depends on the value of $B$. In the limit $T\to0$, the free energy of the isotropic phase is
\begin{equation}
\mathcal{F}_\mathrm{iso}(\phi)= \dfrac{UN}{2}\phi^{2},
\end{equation}
while in the biaxial nematic phase, in which all nematogens are fully aligned with each other, the free energy is
\begin{equation}
    \mathcal{F}_\mathrm{nem}(\phi)=-\dfrac{AN}{4}(3+\Delta^{2})\phi^{2}+\dfrac{3BN}{2}\phi(1-\phi)f(B)+\dfrac{UN}{2}\phi^{2},
\end{equation}
where $f(B)$ takes the value $-1$ if $B>0$ or $(1+\Delta)/2$ if $B<0$. The function $f(B)$ takes different values because different relative orientations between the dipole axes and the first principal nematogen axes are energetically favored, depending on the sign of $B$. In order to minimize the free energy, for $B>0$ ($B<0$), all dipoles have their axes fully aligned with the first (third) principal nematogen axes.

In the interval $0\leq\phi\leq 1$, the above equations imply $\mathcal{F}_\mathrm{nem}(\phi)\leq \mathcal{F}_\mathrm{iso}(\phi)$, with the equality valid only at $\phi=0$. Therefore, the isotropic phase is stable for $\phi=0$ only. However, $\mathcal{F}^{\prime\prime}_\mathrm{nem}(\phi)<0$ for
\begin{equation}
-\frac{2\left(2A -U \right)}{3(1+\Delta)}<B\leq 0\quad\mathrm{or}\quad
0\leq B\leq \frac{2A-U}{3},
\end{equation}
so that, under these conditions, a $\phi=0$ isotropic phase coexists with a $\phi=1$ nematic phase. Otherwise, the nematic phase is stable for $0<\phi\leq 1$.

\section{High-concentration expansion}\label{app:C}
In the limit of high concentration of nematogens we have $\phi=1-\phi_{\mathrm{NP}}$, where $\phi_{\mathrm{NP}}\ll1$ is the concentration of nanoparticles. In this limit, the chemical potential $\mu\to\infty$ and then the fugacity $z\equiv e^{-\beta\mu}\ll1$. If we introduce the scalar functions
\begin{equation}
    a(\bm{\mathrm{Q}})=\sum_{\{\bm{\Theta}\}}e^{\beta B\bm{\mathrm{Q}}\bm{:}\bm{\Theta}},\quad  b(\bm{\mathrm{Q}},\bm{\mathrm{D}})=\sum_{\{\bm{\Omega}\}}e^{\beta(A\bm{\mathrm{Q}}+B\bm{\mathrm{D}})\bm{:}\bm{\Omega}},
\end{equation}
and the tensor functions
\begin{equation}
    c(\bm{\mathrm{Q}},\bm{\mathrm{D}})=\dfrac{1}{\beta A}\cdot\dfrac{\partial}{\partial\bm{\mathrm{Q}}} b(\bm{\mathrm{Q}},\bm{\mathrm{D}}),\quad d(\bm{\mathrm{Q}})=\dfrac{1}{\beta B}\cdot\dfrac{\partial}{\partial\bm{\mathrm{Q}}} a(\bm{\mathrm{Q}}),
\end{equation}
the mean-field equation related to the concentration can be written as
\begin{equation}\label{eq:aphi}
    \phi_{\mathrm{NP}}=1-\left[1+z\dfrac{a(\bm{\mathrm{Q}})}{b(\bm{\mathrm{Q}},\bm{\mathrm{D}})}e^{\beta U(1-\phi_{\mathrm{NP}})}\right]^{-1}\approx z\dfrac{a(\bm{\mathrm{Q}})}{b(\bm{\mathrm{Q}},\bm{\mathrm{D}})}e^{\beta U},
\end{equation}
where the second-rank tensors associated with the nematogens and with the dipolar NPs are $\bm{\mathrm{Q}}=\bm{\mathrm{Q}}^{(0)}+\delta\bm{\mathrm{Q}}$ and $\bm{\mathrm{D}}=\delta\bm{\mathrm{D}}$, respectively, and $\delta\bm{\mathrm{Q}}$ and $\delta\bm{\mathrm{D}}$ are small tensors~\footnote{We say that a second-rank tensor $\bm{\mathrm{G}}$ is small if its entries $g_{ij}$ are small quantities for all pairs $ij$.}. The MF equations related to the order parameters of nematogens and of dipolar nanoparticles are
\begin{equation}
\bm{\mathrm{Q}}^{(0)}+\delta\bm{\mathrm{Q}}=\left[1+z\dfrac{a(\bm{\mathrm{Q}})}{b(\bm{\mathrm{Q}},\bm{\mathrm{D}})}e^{\beta U(1-\phi_{\mathrm{NP}})}\right]^{-1}\dfrac{c(\bm{\mathrm{Q}},\bm{\mathrm{D}})}{b(\bm{\mathrm{Q}},\bm{\mathrm{D}})}=(1-\phi_{\mathrm{NP}})\dfrac{c(\bm{\mathrm{Q}},\bm{\mathrm{D}})}{b(\bm{\mathrm{Q}},\bm{\mathrm{D}})},
\end{equation}
\begin{equation}\label{eq:ad}
    \delta\bm{\mathrm{D}}=ze^{\beta U(1-\phi_{\mathrm{NP}})}\left[1+z\dfrac{a(\bm{\mathrm{Q}})}{b(\bm{\mathrm{Q}},\bm{\mathrm{D}})}e^{\beta U(1-\phi_{\mathrm{NP}})}\right]^{-1}\dfrac{d(\bm{\mathrm{Q}})}{b(\bm{\mathrm{Q}},\bm{\mathrm{D}})}\approx ze^{\beta U}\dfrac{d(\bm{\mathrm{Q}})}{b(\bm{\mathrm{Q}},\bm{\mathrm{D}})},
\end{equation}
and substituting Eq.\,(\ref{eq:aphi}) into Eq.\,(\ref{eq:ad}) we obtain
\begin{equation}
    \delta\bm{\mathrm{D}}\approx\phi_{\mathrm{NP}}\dfrac{d(\bm{\mathrm{Q}})}{a(\bm{\mathrm{Q}})}.
\end{equation}
Notice that
\begin{equation}
    \dfrac{a(\bm{\mathrm{Q}})}{b(\bm{\mathrm{Q}},\bm{\mathrm{D}})}=\dfrac{a^{(0)}(\bm{\mathrm{Q}})+\delta a(\bm{\mathrm{Q}})}{b^{(0)}(\bm{\mathrm{Q}},\bm{\mathrm{D}})+\delta b(\bm{\mathrm{Q}},\bm{\mathrm{D}})}\approx\dfrac{a^{(0)}(\bm{\mathrm{Q}})}{b^{(0)}(\bm{\mathrm{Q}},\bm{\mathrm{D}})}\left[1+\dfrac{\delta a(\bm{\mathrm{Q}})}{a^{(0)}(\bm{\mathrm{Q}})}-\dfrac{\delta b(\bm{\mathrm{Q}}, \bm{\mathrm{D}})}{b^{(0)}\left(\bm{\mathrm{Q}}, \bm{\mathrm{D}}\right)}\right],
\end{equation}
\begin{equation}
    \dfrac{d(\bm{\mathrm{Q}})}{a(\bm{\mathrm{Q}})}=\dfrac{d^{(0)}(\bm{\mathrm{Q}})+\delta d(\bm{\mathrm{Q}})}{a^{(0)}(\bm{\mathrm{Q}})+\delta a(\bm{\mathrm{Q}})}\approx
    \left\{\bm{\mathrm{I}}+\delta d(\bm{\mathrm{Q}})[d^{(0)}(\bm{\mathrm{Q}})]^{-1}-\dfrac{\delta a(\bm{\mathrm{Q}})}{a^{(0)}(\bm{\mathrm{Q}})}\right\}\dfrac{d^{(0)}(\bm{\mathrm{Q}})}{a^{(0)}(\bm{\mathrm{Q}})},
\end{equation}
and
\begin{equation}
    \dfrac{c(\bm{\mathrm{Q}},\bm{\mathrm{D}})}{b(\bm{\mathrm{Q}},\bm{\mathrm{D}})}=\dfrac{c^{(0)}(\bm{\mathrm{Q}},\bm{\mathrm{D}})+\delta c(\bm{\mathrm{Q}},\bm{\mathrm{D}})}{b^{(0)}(\bm{\mathrm{Q}},\bm{\mathrm{D}})+\delta b(\bm{\mathrm{Q}},\bm{\mathrm{D}})}\approx
    \left\{\bm{\mathrm{I}}+\delta c(\bm{\mathrm{Q}},\bm{\mathrm{D}})[c^{(0)}(\bm{\mathrm{Q}},\bm{\mathrm{D}})]^{-1}-\dfrac{\delta b(\bm{\mathrm{Q}}, \bm{\mathrm{D}})}{b^{(0)}\left(\bm{\mathrm{Q}}, \bm{\mathrm{D}}\right)}\right\}\dfrac{c^{(0)}(\bm{\mathrm{Q}},\bm{\mathrm{D}})}{b^{(0)}(\bm{\mathrm{Q}},\bm{\mathrm{D}})},
\end{equation}
with
\begin{equation}
    a^{(0)}(\bm{\mathrm{Q}})=\sum_{\{\bm{\Theta}\}}e^{\beta^{(0)} B\bm{\mathrm{Q}}^{(0)}\bm{:}\bm{\Theta}},\quad  b^{(0)}(\bm{\mathrm{Q}},\bm{\mathrm{D}})=\sum_{\{\bm{\Omega}\}}e^{\beta^{(0)}A\bm{\mathrm{Q}}^{(0)}\bm{:}\bm{\Omega}},
\end{equation}
\begin{equation}
    c^{(0)}(\bm{\mathrm{Q}},\bm{\mathrm{D}})=\sum_{\{\bm{\Omega}\}}e^{\beta^{(0)}A\bm{\mathrm{Q}}^{(0)}\bm{:}\bm{\Omega}}\cdot\bm{\Omega},\quad  d^{(0)}(\bm{\mathrm{Q}})=\sum_{\{\bm{\Theta}\}}e^{\beta^{(0)} B\bm{\mathrm{Q}}^{(0)}\bm{:}\bm{\Theta}}\cdot\bm{\Theta}.
\end{equation}

In the limit of a pure nematogen host ($\phi=1$), we have $\bm{\mathrm{Q}}^{(0)}\equiv\langle\bm{\Omega}\rangle_{0}= c^{(0)}(\bm{\mathrm{Q}},\bm{\mathrm{D}})/b^{(0)}(\bm{\mathrm{Q}},\bm{\mathrm{D}})$, so that the MF equations take the form
\begin{equation}
   \bm{\mathrm{Q}}^{(0)}+ \delta\bm{\mathrm{Q}}=(1-\phi_{\mathrm{NP}})\left\{\bm{\mathrm{I}}+\delta c(\bm{\mathrm{Q}},\bm{\mathrm{D}})[c^{(0)}(\bm{\mathrm{Q}},\bm{\mathrm{D}})]^{-1}-\dfrac{\delta b(\bm{\mathrm{Q}}, \bm{\mathrm{D}})}{b^{(0)}\left(\bm{\mathrm{Q}}, \bm{\mathrm{D}}\right)}\right\}\bm{\mathrm{Q}}^{(0)}.
\end{equation}
Using $[c^{(0)}(\bm{\mathrm{Q}},\bm{\mathrm{D}})]^{-1}\cdot\bm{\mathrm{Q}}^{(0)}=\bm{\mathrm{I}}/b^{(0)}\left(\bm{\mathrm{Q}}, \bm{\mathrm{D}}\right)$ and ignoring terms of order $\phi_{NP}\delta c(\bm{\mathrm{Q}},\bm{\mathrm{D}})$ and $\phi_{NP}\delta b(\bm{\mathrm{Q}},\bm{\mathrm{D}})$, we obtain
\begin{equation}\label{eq:klkl}
    \delta\bm{\mathrm{Q}}=\dfrac{\delta c(\bm{\mathrm{Q}},\bm{\mathrm{D}})}{b^{(0)}(\bm{\mathrm{Q}},\bm{\mathrm{D}})}-\dfrac{\delta b(\bm{\mathrm{Q}}, \bm{\mathrm{D}})}{b^{(0)}(\bm{\mathrm{Q}}, \bm{\mathrm{D}})}\bm{\mathrm{Q}}^{(0)} - \phi_{\mathrm{NP}}\bm{\mathrm{Q}}^{(0)}.
\end{equation}
Note that
\begin{equation}\label{eq:dbdb}
    \delta b(\bm{\mathrm{Q}}, \bm{\mathrm{D}})=\alpha_{\bm{\mathrm{Q}}}\bm{:}\delta\bm{\mathrm{Q}}+\alpha_{\bm{\mathrm{D}}}\bm{:}\delta\bm{\mathrm{D}}+\alpha_{T}\delta T,
\end{equation}
with
\begin{equation}\label{eq:aq}
    \alpha_{\bm{\mathrm{Q}}}=\beta^{(0)}B\sum_{\{\bm{\Omega}\}}e^{\beta^{(0)}A\bm{\mathrm{Q}}^{(0)}\bm{:}\bm{\Omega}}\cdot\bm{\Omega}=\beta^{(0)}Ac^{(0)}(\bm{\mathrm{Q}},\bm{\mathrm{D}}),
\end{equation}
\begin{equation}\label{eq:aed}
    \alpha_{\bm{\mathrm{D}}}=\beta^{(0)}A\sum_{\{\bm{\Omega}\}}e^{\beta^{(0)}A\bm{\mathrm{Q}}^{(0)}\bm{:}\bm{\Omega}}\cdot\bm{\Omega}=\beta^{(0)}Bc^{(0)}(\bm{\mathrm{Q}},\bm{\mathrm{D}}),
\end{equation}
\begin{equation}\label{eq:at}
\begin{split}
    \alpha_{T}&=-(\beta^{(0)})^{2}A\sum_{\{\bm{\Omega}\}}e^{\beta^{(0)}A\bm{\mathrm{Q}}^{(0)}\bm{:}\bm{\Omega}}(\bm{\mathrm{Q}}^{(0)}\bm{:}\bm{\Omega})\\
    &=-(\beta^{(0)})^{2}A\bm{\mathrm{Q}}^{(0)}\bm{:}\sum_{\{\bm{\Omega}\}}e^{\beta^{(0)}A\bm{\mathrm{Q}}^{(0)}\bm{:}\bm{\Omega}}\cdot\bm{\Omega}=-(\beta^{(0)})^{2}A\bm{\mathrm{Q}}^{(0)}\bm{:}c^{(0)}(\bm{\mathrm{Q}},\bm{\mathrm{D}}).
\end{split}
\end{equation}
Then we can insert Eqs.\,(\ref{eq:aq})-(\ref{eq:at}) into the second term on the right-hand side of Eq.\,(\ref{eq:klkl}) and obtain
\begin{equation}\label{eq:ded}
     \dfrac{\delta b(\bm{\mathrm{Q}}, \bm{\mathrm{D}})}{ b(\bm{\mathrm{Q}}, \bm{\mathrm{D}})}=\beta^{(0)}A\bm{\mathrm{Q}}^{(0)}\bm{:}\delta\bm{\mathrm{Q}}+\beta^{(0)}B\bm{\mathrm{Q}}^{(0)}\bm{:}\delta\bm{\mathrm{D}}-(\beta^{(0)})^{2}A\norm{\bm{\mathrm{Q}}^{(0)}}^{2}\delta T.
\end{equation}
On the other hand, we have
\begin{equation}\label{eq:ert}
    \delta c(\bm{\mathrm{Q}}, \bm{\mathrm{D}})=\dfrac{1}{A}\delta\left[\dfrac{1}{\beta}\cdot\dfrac{\partial}{\partial\bm{\mathrm{Q}}}b(\bm{\mathrm{Q}}, \bm{\mathrm{D}})\right]=\dfrac{1}{A}\delta Tc^{(0)}(\bm{\mathrm{Q}}, \bm{\mathrm{D}})+\dfrac{1}{A\beta^{(0)}}\delta\left[\dfrac{\partial}{\partial\bm{\mathrm{Q}}}b(\bm{\mathrm{Q}}, \bm{\mathrm{D}})\right]
\end{equation}
and
\begin{equation}
    \delta\left[\dfrac{\partial}{\partial\bm{\mathrm{Q}}}b(\bm{\mathrm{Q}}, \bm{\mathrm{D}})\right]=\bm{\mathrm{B}}_{\bm{\mathrm{Q}}}\bm{\cdot}\delta\bm{\mathrm{Q}}+\bm{\mathrm{B}}_{\bm{\mathrm{D}}}\bm{\cdot}\delta\bm{\mathrm{D}}+b_{T}\delta T\bm{\mathrm{I}},
\end{equation}
with
\begin{equation}
    \bm{\mathrm{B}}_{\bm{\mathrm{Q}}}=\beta^{(0)}A\sum_{\{\bm{\Omega}\}}e^{\beta^{(0)}A\bm{\mathrm{Q}}^{(0)}\bm{:}\bm{\Omega}}\cdot\bm{\Omega}^{2},
\end{equation}
\begin{equation}
    \bm{\mathrm{B}}_{\bm{\mathrm{D}}}=\beta^{(0)}B\sum_{\{\bm{\Omega}\}}e^{\beta^{(0)}A\bm{\mathrm{Q}}^{(0)}\bm{:}\bm{\Omega}}\cdot\bm{\Omega}^{2},
\end{equation}
\begin{equation}
    b_{T}=-(\beta^{(0)})^{2}A\sum_{\{\bm{\Omega}\}}(\bm{\mathrm{Q}}^{(0)}\bm{:}\bm{\Omega})e^{\beta^{(0)}A\bm{\mathrm{Q}}^{(0)}\bm{:}\bm{\Omega}}\cdot\bm{\Omega}.
\end{equation}
Using Eq.\,(\ref{eq:ert}), we obtain
\begin{equation}
    \dfrac{ \delta c(\bm{\mathrm{Q}}, \bm{\mathrm{D}})}{ b(\bm{\mathrm{Q}}, \bm{\mathrm{D}})}=\dfrac{\delta T}{A}\bm{\mathrm{Q}}^{(0)}+\langle\bm{\Omega}^{2}\rangle_{0}\delta\bm{\mathrm{Q}}+\kappa\langle\bm{\Omega}^{2}\rangle_{0}\delta\bm{\mathrm{D}}+\beta^{(0)}\delta T\langle(\bm{\mathrm{Q}}^{(0)}\bm{:}\bm{\Omega})\cdot\bm{\Omega}\rangle_{0},
\end{equation}
\begin{equation}
    \delta\bm{\mathrm{D}}=\phi_{\mathrm{NP}}\dfrac{d^{(0)}(\bm{\mathrm{Q}})}{a^{(0)}(\bm{\mathrm{Q}})}.
\end{equation}

\subsection{The \texorpdfstring{$\mathrm{ISO}$-$\mathrm{N_{U}}$}{} first-order transition} 
By substituting the mean-field equations (\ref{eq:mf1}) to (\ref{eq:mf3}) into Eq.\,(\ref{eq:LdG}), the Landau--de Gennes free-energy functional can be written as
\begin{equation}
    {\psi}(\bm{\mathrm{Q}}, \bm{\mathrm{D}}, \phi)=\dfrac{A}{2}\norm{\bm{\mathrm{Q}}}^{2}+B\bm{\mathrm{Q}}\bm{:}\bm{\mathrm{D}}-\dfrac{U}{2}\phi^{2}-\dfrac{1}{\beta}\ln\left[a(\bm{\mathrm{Q}})+\dfrac{e^{-\beta U\phi}}{z}b(\bm{\mathrm{Q}}, \bm{\mathrm{D}})\right].
\end{equation}
In the limit $z\to0$, we can write
\begin{equation}
\begin{split}
    &\dfrac{1}{\beta}\ln\left[a(\bm{\mathrm{Q}})+\dfrac{e^{-\beta U\phi}}{z}b(\bm{\mathrm{Q}}, \bm{\mathrm{D}})\right]=\dfrac{1}{\beta^{(0)}}\left[\dfrac{a^{(0)}(\bm{\mathrm{Q}})}{b^{(0)}(\bm{\mathrm{Q}}, \bm{\mathrm{D}})}e^{\beta^{(0)}U}z-\ln(z)\right]\\
    &\qquad-U\phi+\dfrac{1}{\beta^{(0)}}\ln[b^{(0)}(\bm{\mathrm{Q}}, \bm{\mathrm{D}})]+\delta\left\{\dfrac{1}{\beta}\ln[b(\bm{\mathrm{Q}}, \bm{\mathrm{D}})]\right\}+\mathcal{O}(z).
\end{split}
\end{equation}
Notice that
\begin{equation}
    \delta\left\{\dfrac{1}{\beta}\ln[b(\bm{\mathrm{Q}}, \bm{\mathrm{D}})]\right\}=\ln[b^{(0)}(\bm{\mathrm{Q}}, \bm{\mathrm{D}})]\delta T + T   \delta\left\{\ln[b(\bm{\mathrm{Q}}, \bm{\mathrm{D}})]\right\},
\end{equation}
with
\begin{equation}
     \delta\left\{\ln[b(\bm{\mathrm{Q}}, \bm{\mathrm{D}})]\right\}=\dfrac{\delta\{b(\bm{\mathrm{Q}}, \bm{\mathrm{D}})\}}{b(\bm{\mathrm{Q}}, \bm{\mathrm{D}})}.
\end{equation}
Using $\delta\{\norm{\bm{\mathrm{Q}}}^{2}\}=2\bm{\mathrm{Q}}^{(0)}\bm{:}\delta\bm{\mathrm{Q}}$ and $\delta\{\bm{\mathrm{Q}}\bm{:}\bm{\mathrm{D}}\}=\bm{\mathrm{Q}}^{(0)}\bm{:}\delta\bm{\mathrm{D}}$ in Eq.\,(\ref{eq:ded}), we obtain
\begin{equation}
\begin{split}
    &{\psi}(\bm{\mathrm{Q}}, \bm{\mathrm{D}}, \phi)=\dfrac{A}{2}\norm{\bm{\mathrm{Q}}^{(0)}}^{2}+A\bm{\mathrm{Q}}^{(0)}\bm{:}\delta\bm{\mathrm{Q}} +B\bm{\mathrm{Q}}^{(0)}\bm{:}\delta\bm{\mathrm{D}}-\dfrac{U}{2}-\dfrac{1}{\beta^{(0)}}\ln[b^{(0)}(\bm{\mathrm{Q}}, 
    \bm{\mathrm{D}})]\\
    &\qquad\qquad-\ln[b^{(0)}(\bm{\mathrm{Q}}, \bm{\mathrm{D}})]\delta T -A\bm{\mathrm{Q}}^{(0)}\bm{:}\delta\bm{\mathrm{Q}}-B\bm{\mathrm{Q}}^{(0)}\bm{:}\delta\bm{\mathrm{D}}+\beta^{(0)}A\norm{\bm{\mathrm{Q}}^{(0)}}^{2}\delta T\\
    &\qquad\qquad-\dfrac{1}{\beta^{(0)}}\left[\dfrac{a^{(0)}(\bm{\mathrm{Q}})}{b^{(0)}(\bm{\mathrm{Q}}, \bm{\mathrm{D}})}e^{\beta^{(0)}U}z-\ln(z)\right]+\mathcal{O}(z),
    \end{split}
\end{equation}
which is equivalent to
\begin{equation}\label{eq:nematicc}
\begin{split}
    {\psi}(\bm{\mathrm{Q}}, \bm{\mathrm{D}}, \phi)&=\dfrac{A}{2}\norm{\bm{\mathrm{Q}}^{(0)}}^{2}-\dfrac{U}{2}-\dfrac{1}{\beta^{(0)}}\ln[b^{(0)}(\bm{\mathrm{Q}}, 
    \bm{\mathrm{D}})]\\
    &
    -\ln[b^{(0)}(\bm{\mathrm{Q}}, \bm{\mathrm{D}})]\delta T +\beta^{(0)}A\norm{\bm{\mathrm{Q}}^{(0)}}^{2}\delta T\\
    &-\dfrac{1}{\beta^{(0)}}\left[\dfrac{a^{(0)}(\bm{\mathrm{Q}})}{b^{(0)}(\bm{\mathrm{Q}}, \bm{\mathrm{D}})}e^{\beta^{(0)}U}z-\ln(z)\right]+\mathcal{O}(z).
\end{split}
\end{equation}

We know that for the $\mathrm{ISO}$-$\mathrm{N^{+}_U}$ first-order transition $\psi(\bm{0},\bm{0},\phi_{\mathrm{I}})= \psi(\bm{\mathrm{Q}}_{\mathrm{U}}, \bm{\mathrm{D}}_{\mathrm{U}}, \phi_{\mathrm{U}})$, then 
for the $\mathrm{ISO}$ phase we have that $a^{(0)}(\bm{\mathrm{Q}})=a^{(0)}(\bm{0})=6$ and $b^{(0)}(\bm{\mathrm{Q}}, \bm{\mathrm{D}})=b^{(0)}(\bm{0}, \bm{0})=6$, so that
\begin{equation}\label{eq:isotr}
    {\psi}(\bm{0},\bm{0},\phi_{\mathrm{I}})=-\dfrac{U}{2}-\dfrac{1}{\beta^{(0)}}\ln(6)-\ln(6)\delta T-\dfrac{1}{\beta^{(0)}}\left[e^{\beta^{(0)}U}z-\ln(z)\right]+\mathcal{O}(z).
\end{equation}
A comparison between Eq.\,(\ref{eq:isotr}) and Eq.\,(\ref{eq:nematicc}) yields
\begin{equation}
\begin{split}
   & -\dfrac{1}{\beta^{(0)}}\ln(6)-\ln(6)\delta T-\dfrac{1}{\beta^{(0)}}\left[e^{\beta^{(0)}U}z\right]=\dfrac{A}{2}\norm{\bm{\mathrm{Q}}_{\mathrm{U}}^{(0)}}^{2}-\dfrac{1}{\beta^{(0)}}\ln[b^{(0)}(\bm{\mathrm{Q}}_{\mathrm{U}}, 
    \bm{\mathrm{D}}_{\mathrm{U}})]\\
    &  -\ln[b^{(0)}(\bm{\mathrm{Q}}_{\mathrm{U}}, 
    \bm{\mathrm{D}}_{\mathrm{U}})]\delta T +\beta^{(0)}A\norm{\bm{\mathrm{Q}}^{(0)}_{\mathrm{U}}}^{2}\delta T-\dfrac{1}{\beta^{(0)}}\left[\dfrac{a^{(0)}(\bm{\mathrm{Q}})}{b^{(0)}(\bm{\mathrm{Q}}, \bm{\mathrm{D}})}e^{\beta^{(0)}U}z\right],
    \end{split}
\end{equation}
but for the pure nematic we have
\begin{equation}
    \dfrac{A}{2}\norm{\bm{\mathrm{Q}}_{\mathrm{U}}^{(0)}}^{2}-\dfrac{1}{\beta^{(0)}}\ln[b^{(0)}(\bm{\mathrm{Q}}_{\mathrm{U}}, 
    \bm{\mathrm{D}}_{\mathrm{U}})]=-\dfrac{1}{\beta^{(0)}}\ln(6),
\end{equation}
so that
\begin{equation}
       - \beta^{(0)}\dfrac{A}{2}\norm{\bm{\mathrm{Q}}^{(0)}_{\mathrm{U}}}^{2}\delta T=
     \dfrac{e^{\beta^{(0)}U}}{\beta^{(0)}}\left[1-\dfrac{a^{(0)}(\bm{\mathrm{Q}}_{\mathrm{U}})}{b^{(0)}(\bm{\mathrm{Q}}_{\mathrm{U}}, \bm{\mathrm{D}}_{\mathrm{U}})}\right]z.
\end{equation}
Using Eq.\,(\ref{eq:aphi}), we have $ze^{\beta^{(0)}U}=\phi_{\mathrm{NP}}b^{(0)}(\bm{\mathrm{Q}}_{\mathrm{U}}, \bm{\mathrm{D}}_{\mathrm{U}})/a^{(0)}(\bm{\mathrm{Q}}_{\mathrm{U}})$, then
\begin{equation}
  \dfrac{\,dT_{\mathrm{I}-\mathrm{U}}}{\,d\phi_{\mathrm{NP}}}= \dfrac{\delta T_{\mathrm{I}-\mathrm{U}}}{\phi_{\mathrm{NP}}} =  \dfrac{2}{A}\left[1-\dfrac{b^{(0)}(\bm{\mathrm{Q}}_{\mathrm{U}}, \bm{\mathrm{D}}_{\mathrm{U}})}{a^{(0)}(\bm{\mathrm{Q}}_{\mathrm{U}})}\right]\left(\dfrac{T^{(0)}_{\mathrm{I}-\mathrm{U}}}{\norm{\bm{\mathrm{Q}}^{(0)}_{\mathrm{U}}}}\right)^{2}.
\end{equation}
Notice that for $a^{(0)}(\bm{\mathrm{Q}}_{\mathrm{U}})=b^{(0)}(\bm{\mathrm{Q}}_{\mathrm{U}}, \bm{\mathrm{D}}_{\mathrm{U}})$ this derivative is zero.

\subsection{The \texorpdfstring{$\mathrm{N_{U}}$-$\mathrm{N_{B}}$}{} second-order transition}
The perturbative calculations for the $\mathrm{N_U}$-$\mathrm{N_B}$ second-order transition are more complicated from an algebraic point of view. The conditions for the second-order transition are the MF equations supplemented by $\,d^{2}\psi/\,d\eta^{2}=0$. In this case we solve the problem computationally using Wolfram Mathematica; see~\cite{william2023second}.

\bibliography{apssamp}

\end{document}